\documentclass{article}
\usepackage{graphicx} 
\usepackage{caption} 
\usepackage{amsmath}
\usepackage{amssymb}
\usepackage{subcaption}
\usepackage{bm}

\newcommand{\e}{\mathbb{E}}

\newcommand{\Var}{\mathrm{Var}}
\newcommand{\Cov}{\mathrm{Cov}}
\newcommand{\tran}{\mathsf{T}}
\usepackage{hyperref} 
\usepackage{booktabs}
\usepackage{authblk}
\usepackage{multirow}
\usepackage{comment}
\usepackage{xcolor}
\usepackage{pifont}
\hypersetup{
    colorlinks=true,
    linkcolor=blue,
    urlcolor=red,
    linktoc=all,
    citecolor=green
           }
\usepackage[natbib, style=authoryear-comp, maxcitenames=2, uniquelist=false]{biblatex}
\usepackage[margin=1in]{geometry}
\usepackage[title]{appendix}
\usepackage[colorinlistoftodos]{todonotes}

\let\cite\citep

\DeclareMathOperator*{\argmin}{arg\,min}

\definecolor{green2}{rgb}{0.0, 0.70, 0.0}

\newcommand{\dan}[1]{{\color{orange}[DK: {#1}]}}

\begin{document}
\title{Regression coefficient estimation from remote sensing maps}

\author[1,2]{Kerri Lu\thanks{Email: \{\href{mailto:kerrilu@mit.edu}{kerrilu}, \href{mailto:dkluger@mit.edu}{dkluger}, \href{mailto:stephenbates@mit.edu}{stephenbates}, \href{mailto:sherwang@mit.edu}{sherwang}\}@mit.edu}}
\author[4]{Dan M. Kluger$^{*}$}
\author[1,2]{Stephen Bates$^{*}$}
\author[1,3,4]{Sherrie Wang$^{*}$}
\affil[1]{Laboratory for Information and Decision Systems, MIT}
\affil[2]{Department of Electrical Engineering and Computer Science, MIT}
\affil[3]{Department of Mechanical Engineering, MIT}
\affil[4]{Institute for Data, Systems, and Society, MIT}
\date{}
\maketitle
\begin{abstract}
Regressions are commonly used in environmental science and economics to identify causal or associative relationships between variables. In these settings, remote sensing-derived map products increasingly serve as sources of variables, enabling estimation of effects such as the impact of conservation zones on deforestation. However, the quality of map products varies, and --- because maps are outputs of complex machine learning algorithms that take in a variety of remotely sensed variables as inputs --- errors are difficult to characterize. Thus, population-level estimators from such maps may be biased. In this paper, we apply prediction-powered inference (PPI) to estimate regression coefficients relating a response variable and covariates to each other. PPI is a method that estimates parameters of interest by using a small amount of randomly sampled ground truth data to correct for bias in large-scale remote sensing map products. Applying PPI across multiple remote sensing use cases in regression coefficient estimation, we find that it results in estimates that are (1) more reliable than using the map product as if it were 100\% accurate and (2) have lower uncertainty than using only the ground truth sample data and ignoring the map product. Empirically, we observe effective sample size increases of up to 17-fold using PPI compared to only using ground truth data. This is the first work to 
estimate remote sensing regression coefficients without assumptions on the structure of map product errors. 
Data and code are available at \url{https://github.com/Earth-Intelligence-Lab/uncertainty-quantification}.
\end{abstract}

\section{Introduction}\label{sec:introduction}
Remote sensing maps are widely used to obtain estimates for environmental quantities, such as cropland expansion over time or the effect of protected areas on deforestation \cite{hansen2013high, baccini2012estimated, potapov2022global, wang2019great, deines2019satellites}. Map products created by training machine learning models on remotely sensed data allow researchers to access predictions for a variable of interest at a high resolution over large regions. 

However, the quality of map products varies, and resulting population-level estimators of quantities of interest may be biased \cite{jain2020benefits, bastin2017extent, venter2024uncertainty}. For example, \citet{bastin2017extent} showed that human labeling of high-resolution imagery results in an estimate of global forest extent in dryland biomes that is 40\% to 47\% higher than previous estimates from remotely sensed forest cover map products. 
Moreover, different map products may conflict with each other and result in significantly different estimates over the same region. For example, \citet{alix2023remotely} found that two commonly used satellite-based remote sensing maps of forest cover in Mexico differ significantly from each other, implying that there is misclassification error in at least one of the algorithms used to categorize pixels as forest or non-forest. Another recent study \cite{venter2022global} of three satellite-based global land cover/land use map products found that the maps are biased toward different land cover classes (e.g., one map overestimates shrub cover and another overestimates grass cover relative to the other maps). While some of these discrepancies may be due to differences in land cover class definitions between maps, they also point to the possibility of systematic error in the map products.

Errors in map products are often difficult to characterize due to the complexity of input data and lack of interpretability in the ML models used to generate the maps. Examples of systematic errors in map products trained on satellite data include ``nonrandom misclassification, saturation effects, atmospheric effects, and cloud cover," all of which may lead to biased inference \cite{jain2020benefits}. This has prevented environmental scientists, economists, and policymakers from using remote sensing maps in their analyses. The \citet{world2018mainstreaming} has cited uncertainty in remote sensing maps as a barrier to their use in water resources management 
, and the \citet{united2017earth} wrote that the 
``shift in paradigm from traditional statistical methods (e.g., counting, measuring by humans) towards estimation from sensors (with associated aspects of validity, approximation, uncertainty)...
will require convincing, statistically sound results." Correctly accounting for map errors, then, is crucial for maximizing the potential of remote sensing data to expand scientific knowledge and advance practical operations.

For those who currently reject estimation from remote sensing as too unreliable, the alternative is to use ground truth observations for the variable of interest.
However, ground truth data are often sparse because they are expensive to collect, as this may require fieldwork or manual labelling of remote sensing imagery. Few ground truth data points result in greater uncertainty (i.e., wider confidence intervals) in estimates. (Throughout this paper, we will use ``data points" to refer to individual sampling units, such as pixels or grid cells, rather than precise latitude-longitude pairs.)

\begin{figure}[t]
\centering
\includegraphics[width=\textwidth]{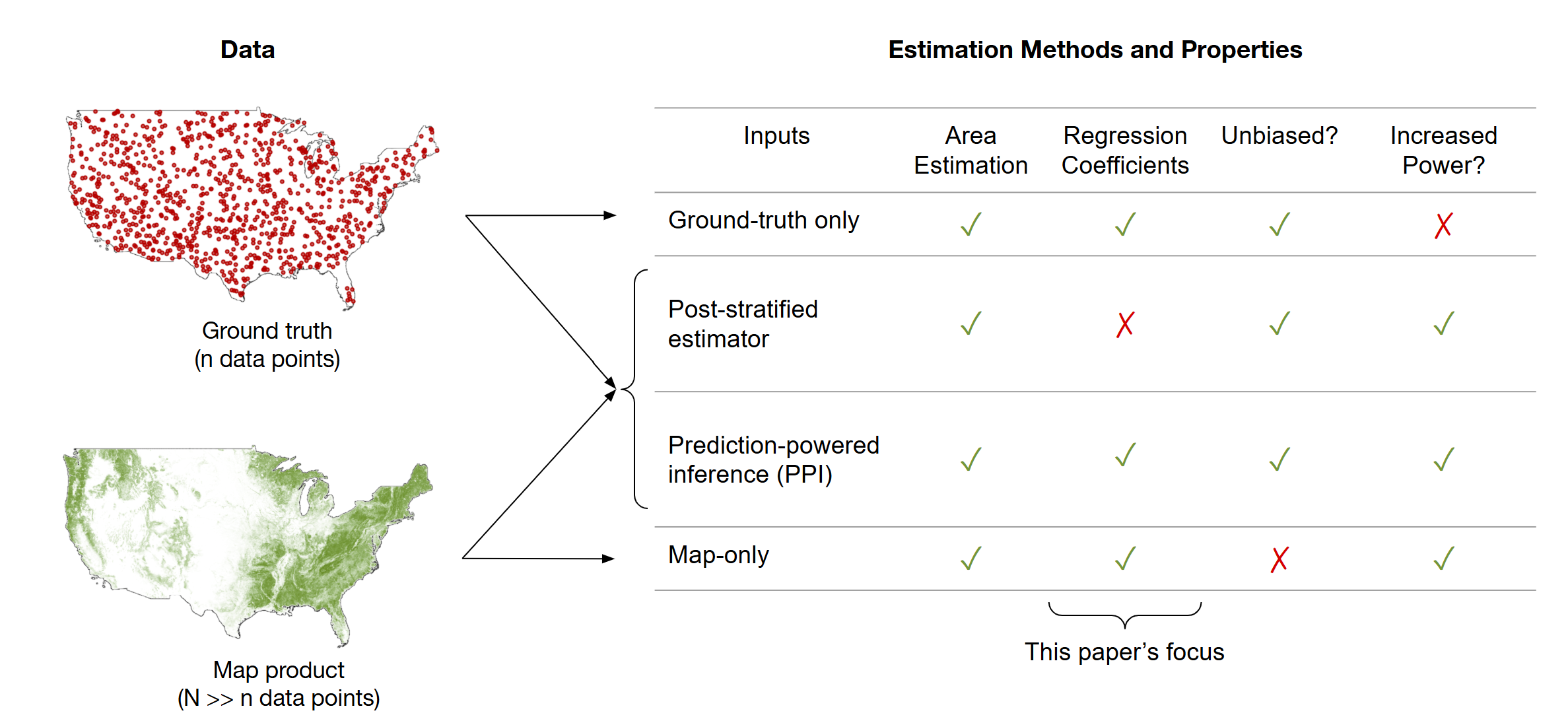}
\caption{\textbf{Overview of estimation methods using ground truth data and remote sensing map products.} The post-stratified estimator (for area estimation) and prediction-powered inference (for regression coefficient estimation) use a small amount of ground truth data along with a map product to produce confidence intervals for the quantity of interest. In this paper, we focus on regression coefficient estimation using PPI. We also compare PPI against GT-only (using only ground truth) and map-only estimators.}\label{summary}
\end{figure}

Another alternative is to use both ground truth data and remote sensing maps to generate estimates, using the ground truth data points to correct for biases that may exist in the map products. This generally reduces uncertainty in the estimates compared to using ground truth alone, since map products contain a large amount of data that, while possibly biased, still give useful information about the true values we wish to measure. 

As an example of this approach, the remote sensing community commonly uses a \textbf{post-stratified estimator} to generate confidence intervals for \textbf{area estimation}. A \textbf{pixel-counting estimator}, which uses only a map product to estimate area of a land cover class, may be biased due to misclassification error in the map product \cite{olofsson2020mitigating}. An unbiased area estimator can be obtained by using observations of reference conditions in locations selected by probability sampling \cite{olofsson2014good, stehman2019key}, but if ground truth observations are sparse, this may result in less precise estimates (wider confidence intervals) than desired. To resolve this, the post-stratified area estimator \cite{card1982using, stehman2013estimating, olofsson2013making} combines a map product with a small simple random sample or systematic sample of ground truth data, using the confusion matrix between land cover ground truth and map product data to correct for bias in the map product. By leveraging the map product, the post-stratified estimator also produces narrower confidence intervals compared to only using ground truth data. When the ground truth data are instead selected via stratified random sampling, a \textbf{stratified estimator} is commonly used \cite{cochran1977sampling, olofsson2013making, olofsson2014good}. (See Appendix \ref{appendix:area-estimation} for more detail on stratified and post-stratified area estimators.) Other area estimators that combine ground truth and map product data, including regression, ratio, and calibration estimators, are described in \citet{gallego2004remote}. 

The remote sensing community also uses map products for \textbf{regression coefficient estimation}: estimating linear or logistic regression coefficients to assess causal or associative relationships between two or more environmental variables. Unlike area estimation with the post-stratified estimator, regression coefficient estimation from remote sensing maps still suffers from bias due to map product errors. An unbiased estimator can be obtained by running a regression on a probability sample of ground truth data \cite{sarndal1992model}, but similarly to the area estimation case, this may result in less precise estimates than desired. 

In this paper, we address this problem by applying \textbf{prediction-powered inference (PPI)} to regression coefficient estimation (Figure \ref{summary}). PPI \cite{angelopoulos2023prediction, angelopoulos2023ppi++,Kluger25GeneralizingPPI} is a recently developed method that computes point estimates and confidence intervals for parameters of interest by using a small randomly selected set (i.e., a probability sample) of ground truth observations ($n$ data points) to calibrate and correct for bias in a much larger set of machine learning-generated predictions ($N$ data points, where $N$ is much larger than $n$). This method also results in more precise estimates (higher statistical power, narrower confidence intervals) than using the ground truth data points alone. We apply PPI to regression use cases where the machine learning predictions are obtained from large-scale remote sensing map products. We compare results with the \textbf{ground truth-only (GT-only) estimator} (using only ground truth data points) and \textbf{map-only estimator} (using only a map product). These estimators are described in detail in Section \ref{section:methods}. 

\begin{figure}[t]
\centering
\includegraphics[width=0.8\textwidth]{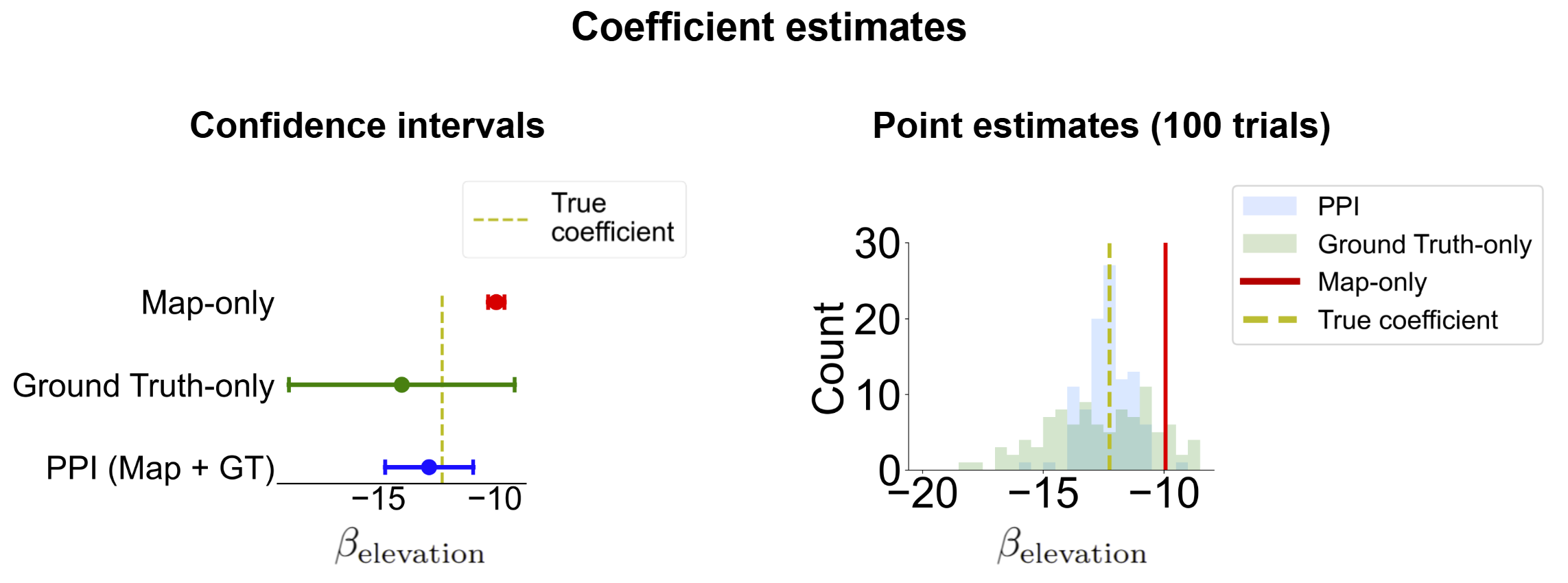}
\caption{\textbf{Motivating example: PPI gives an unbiased regression coefficient estimator with lower variance than using ground truth alone.} We estimate the linear regression coefficient relating US forest cover to elevation (with $N=67968$ remote sensing map data points and $n=500$ ground truth data points). Left: Map-only, ground truth-only, and PPI point estimates and 95\% confidence intervals. Right: Histogram of PPI and GT-only point estimates from repeating the experiment with 100 simple random subsamples of $n=500$ ground truth data points.}\label{motivating-example}
\end{figure}

\paragraph{A motivating example.} 

We give a brief preview of how we use the PPI regression coefficient estimator to correct map product bias with ground-truth data. 

We seek to estimate linear regression coefficients for the association between US forest cover and a few covariates, including elevation. We use $N=67968$ remote sensing map product data points and $n=500$ ground truth data points. See Section \ref{subsection:use-case-datasets} for a complete description of the experimental setup.

We compare the $95\%$ confidence intervals for the elevation coefficient estimates from PPI, GT-only, and map-only methods (Figure \ref{motivating-example}, left panel). The ``GT-only'' estimate using only ground truth data points is unbiased but has the widest confidence interval, failing to take advantage of existing map products. Meanwhile, PPI uses the map product along with the ground truth data points to produce a much narrower confidence interval while still being unbiased. The PPI and GT-only confidence intervals both contain the true coefficient, while the confidence interval estimated from the map product alone does not. That is, if we had used only the map product, we would have underestimated the magnitude of the negative relationship between forest cover and elevation. PPI avoids misleading conclusions by using ground truth data points to correct for bias, and at the same time, it gives narrower confidence intervals than using only the ground truth data points.

We repeat the experiment using the same set of $N=67968$ map product data points and 100 different simple random subsamples of $n=500$ ground truth data points (see Appendix \ref{appendix:datasets-forest-cover} for more detail). A histogram of the 100 PPI and GT-only elevation coefficient point estimates is shown in the right panel of Figure \ref{motivating-example}. As before, we see that the PPI and GT-only estimators are both unbiased relative to the true coefficient, but the PPI estimates are more precise as they have a lower variance than the GT-only estimates. 

Note that we repeat the experiment 100 times for illustrative purposes only; we are not using the 100 PPI and GT-only point estimates to compute the variance of the estimators or to construct the confidence intervals shown in the left panel of Figure \ref{motivating-example}. The three confidence intervals are constructed under a superpopulation-based inference framework (Appendix \ref{appendix:SuperpopulationInferenceJustification}), using a single sample of $n=500$ ground truth data points and $N=67968$ map product data points. The PPI confidence interval is constructed using a bootstrap procedure described in Section \ref{subsection:linear-regression}. The GT-only and map-only confidence intervals are constructed using classical formulas for the variance of linear regression coefficients. 

\paragraph{Our contribution.}
In this paper, we illustrate the use of PPI on four specific regression coefficient estimation tasks associating deforestation, tree cover, housing price, and forest cover to other geospatial variables; see Table~\ref{use-cases}. We then simulate map error to understand its influence on the bias and variance of map-only, GT-only, and PPI estimators. We demonstrate that PPI can be applied when there is error in the remotely sensed response variable, the remotely sensed covariates, or both. Data and code for these use cases are available at \url{https://github.com/Earth-Intelligence-Lab/uncertainty-quantification}.  

\begin{table}[t]
\centering
\scriptsize
\begin{tabular}{p{1.2cm}p{1.4cm}p{2.5cm}p{1.2cm}p{0.6cm}p{3cm}p{2.3cm}p{1cm}}
\toprule
\textbf{Regression type} & \textbf{Response variable} & \textbf{Covariates} & \textbf{Region} & \textbf{Year} & \textbf{Map product(s)} & \textbf{Ground truth} & \textbf{Link to code}\\ \midrule
Logistic & Deforestation & Distance from roads, distance from rivers, elevation, slope & Brazilian Amazon & 2000-2015 & NASA Global Forest Cover Change \cite{townshend2016global} & $n=1386$ \newline from \cite{bullock2020satellite} & \href{https://github.com/Earth-Intelligence-Lab/uncertainty-quantification/blob/main/examples/deforestation%20logistic%20regression.ipynb}{link}\\ \midrule
Linear & Tree cover & Aridity index, \newline elevation, slope & Contiguous USA & 2021 & USFS Tree Canopy Cover \cite{forest2023usfs} & $n=983$ \newline manually labeled images (Sec. \ref{subsection:use-case-datasets}) & \href{https://github.com/Earth-Intelligence-Lab/uncertainty-quantification/blob/main/examples/tree%20cover%20linear%20regression.ipynb}{link}\\ \midrule
Linear & Housing price & Income, nightlights, road length & Contiguous USA & 2010-2021 & MOSAIKS \cite{rolf2021generalizable} & $n=500$ \newline from \cite{rolf2021generalizable} & \href{https://github.com/Earth-Intelligence-Lab/uncertainty-quantification/blob/main/examples/housing%20price%20linear%20regression.ipynb}{link}\\ \midrule
Linear & Forest cover & Elevation, \newline population & Contiguous USA & 2010 & MOSAIKS \cite{rolf2021generalizable} & $n=500$ \newline from \cite{rolf2021generalizable} & \href{https://github.com/Earth-Intelligence-Lab/uncertainty-quantification/blob/main/examples/forest%20cover%20linear%20regression.ipynb}{link}\\ \bottomrule
\end{tabular}
\caption{\textbf{Overview of use cases and datasets.} To illustrate the mechanics and advantages of PPI, we estimate linear regression coefficients for tree cover, housing price, and forest cover in the United States. Code implementations of the use cases are linked in the rightmost column.}\label{use-cases}
\end{table}

PPI is a method that combines ground truth data with map products to produce an unbiased estimator that is more reliable than the map-only estimator and results in confidence intervals with lower uncertainty (i.e., narrower widths) than the GT-only estimator. As we will show, the map-only estimator may be biased. 

To improve the utility of machine learning-generated remote sensing maps for downstream regression applications, we recommend that map \emph{producers} provide a randomly sampled (i.e., via a probability sampling design) holdout ground truth dataset to be used for calibration in PPI alongside their maps, or, if that is not done, for map \emph{users} to generate their own randomly sampled holdout set for calibration, if possible (e.g., from visual inspection of remote sensing data). This holdout set can be obtained by simple random sampling or other probability sampling designs such as unequal probability sampling, stratified random sampling, or cluster random sampling \cite{Kluger25GeneralizingPPI}. 

\paragraph{Related work.} Previous works have developed methods to reduce bias in regression coefficient estimators from remote sensing data. For example, \citet{alix2023remotely} estimates the effect of a conservation program on deforestation in Mexico. They correct for bias in the remotely sensed deforestation binary variable by modelling misclassification probabilities as a parametric function of environmental covariates. However, this approach imposes assumptions about the structure of the misclassification errors and moreover is only developed for settings with binary response variables. Another study \cite{gordon2023remote} estimates logistic regression coefficients for the relationship between roads and forest cover in West Africa. They train an adversarial debiasing model with ground truth data to correct for bias arising from correlations between distance from roads and the measurement error in remotely sensed forest cover; however, they do not provide an approach for constructing confidence intervals. Another study \cite{proctor2023parameter} uses co-located ground truth and remotely sensed data to show that using remote sensing data alone leads to biased regression coefficient estimators for several effect size estimation tasks in the US. To remedy this issue, they propose using a Bayesian linear measurement error model and applying multiple imputation: a statistical method that involves estimating the missing values in a dataset and their uncertainties and fitting the regression on multiple randomly imputed versions of the dataset. Note that this method requires an assumption about the error structure of the remotely-sensed variables and does not have any frequentist coverage guarantees. For example, in a subset of the simulation-based experiments in \citet{proctor2023parameter}, 95\% confidence intervals from multiple imputation only contained the true value of the regression coefficient in fewer than 80\% of the simulations as opposed to the desired $\sim$95\% of the simulations. In contrast to existing approaches in the remote sensing literature, PPI does not make any modelling assumptions about the measurement error structure, and under certain regularity conditions, its confidence intervals are guaranteed to attain the desired coverage \cite{angelopoulos2023prediction,angelopoulos2023ppi++,Kluger25GeneralizingPPI}. 

Another approach to uncertainty quantification is \textit{conformal prediction}~\cite{vovk2005algorithmic, angelopoulos2023conformal}, which generates confidence sets for machine learning predictions at individual data points. For example, \citet{valle2023quantifying} applies conformal prediction to quantify pixel-level uncertainty for a land cover/land use map product in the Brazilian Amazon, outputting a 90\% predictive set of land cover classes for each pixel (i.e., the set contains the true land cover class 90\% of the time). This approach differs from our work because PPI estimates regression coefficients over an entire region, rather than quantifying uncertainty of map estimates at individual data points or pixels.    

Some readers may also be reminded of the regression estimator, which uses a regression to improve precision in mean estimates. We refer to this method as the \textit{regression mean estimator}; it combines a small random sample of ground truth data points with map data or widely measured covariates to estimate the mean of a variable or the area of a land cover class \cite{cochran1977sampling, gallego2004remote}. Several papers use the regression mean estimator to combine satellite data and field data to estimate crop areas \cite{allen1990look, taylor1997regional, li2023development, alonso1991comparing}. Although this estimator involves a regression, it outputs mean or area estimates, while our work focuses on estimating regression coefficients relating different variables. (See Appendix \ref{appendix:regression-estimator} for more details and a description of the notable parallels between the PPI estimator we study and the regression mean estimator.)

Unlike previous works, this paper is the first work to apply prediction-powered inference to remote sensing use cases in regression coefficient estimation. This is also the first work to combine map product and ground truth data to estimate remote sensing regression coefficients without assuming an error structure on the map product predictions. We apply PPI to both real data examples (Section \ref{sec:examples}) and simulated examples (Section \ref{sec:simulations}); the simulations allow us to control various properties of map error and observe their effects on the resulting estimates and confidence intervals.  


\section{Methods: Prediction-Powered Inference for Regression}\label{section:methods}
Regressions are commonly used in the environmental sciences and economics to identify relationships --- associative or causal --- between covariates and a binary response (logistic regression) or between covariates and a continuous response (linear regression). Logistic regression coefficients give the change in the log-odds ratio of a binary response corresponding to a one-unit increase in a covariate, when all other covariates are held constant. Linear regression coefficients give the change in the value of a continuous response corresponding to a one-unit increase in a covariate, when all other covariates are held constant. Thus, we can interpret these coefficients as effect sizes of the covariates. 

Historically, scientists would run regressions using ground truth environmental samples, but regressions on data from remote sensing maps are becoming increasingly common, e.g., \cite{taylor2023fertilizer, deines2023recent, he2022marked, taylor2022wetlands, deines2019satellites, lobell2022globally}. When covariate or response variable values are machine learning predictions from a map product, errors in the machine learning model may result in biased regression coefficients (as we have seen in Figure \ref{motivating-example}). 

In this section, we formally describe how to use the \textbf{prediction-powered inference (PPI)} method to generate regression coefficient estimates and confidence intervals when the response variable $Y$ and/or covariates $X$ are remotely sensed map products. As two baselines, we consider the \textbf{ground truth-only estimator} that only uses ground-truth data (which is unbiased but does not leverage remote sensing maps) and the \textbf{map-only estimator} that only uses remote sensing maps (which may be biased). We then describe the PPI estimator, which is unbiased and uses both sources of data to produce confidence intervals that are guaranteed to attain the desired coverage. Figure~\ref{summary} summarizes the methods. As a technical note, this paper focuses on inference in settings where both the ground truth data and the larger sample from a remote sensing map product are assumed to be selected using probability sampling from a superpopulation, as opposed to a design-based inference framework for reasons discussed in Appendix \ref{appendix:SuperpopulationInferenceJustification}. 

Suppose we have a remote sensing data product with a large number $N$ of predictions $(\hat{X}'_1, \hat{Y}'_1)$, $(\hat{X}'_2, \hat{Y}'_2)$, $\dots$, $(\hat{X}'_N, \hat{Y}'_N)$ where $X \in \mathbb{R}^p$ is a vector of covariates and $Y \in \mathbb{R}$ is the response variable. Suppose that for a small \textbf{calibration set} of $n \ll N$ data points selected via simple random sampling (such that the $n$ data points are i.i.d.), we have ground truth labels $(X_1, Y_1)$, $(X_2, Y_2)$,$\dots$, $(X_n, Y_n)$ with corresponding remote sensing predictions $(\hat{X}_1, \hat{Y}_1), (\hat{X}_2, \hat{Y}_2), \dots, (\hat{X}_n, \hat{Y}_n)$. (Note that since the calibration data points are randomly sampled, they do not necessarily have the same numbering as the set of all extracted remote sensing predictions; for example, $\hat{X}_1$ is not necessarily the same as $\hat{X}'_1$.) In some use cases, ground truth data may be widely available either for the response variable, for a subset of the covariates, or for a subset of both. In these cases, we would use ground truth instead of remotely sensed values for the corresponding components of the $(\hat{X}, \hat{Y})$ and $(\hat{X}', \hat{Y}')$ vectors.

\subsection{Linear regression}\label{subsection:linear-regression} 
In the case of linear regression, we wish to estimate 95\% confidence intervals for the regression coefficients $\beta_*$ giving the best fit to the linear model
\begin{align*} 
    Y = \beta^\tran X  + \varepsilon,
\end{align*} where $\varepsilon$ is mean $0$ and uncorrelated with $X$. (We assume that the first element of the vector $X$ is a constant 1, so that the first regression coefficient $\beta_0$ is an intercept term.)

\paragraph{Ground truth-only estimator.} The ground truth-only (GT-only) estimator uses only the $n$ ground truth calibration data points to compute the regression coefficient estimate
\begin{align*}
    \hat{\beta}_{\text{calib}} = \argmin_{\beta} \sum_{i=1}^n ( Y_i - \beta^\tran X_i)^2.
\end{align*}
However, the standard error of this ground truth coefficient estimate will be large when the number of ground truth observations $n$ is small, which is often the case in environmental applications. As a result, the GT-only 95\% confidence interval will be wide.  

\paragraph{Map-only estimator: Use only the remote sensing map.} The map-only estimator uses only the $N$ map data points to compute the regression coefficient estimate
\begin{align*}
    \hat{\gamma}_{\text{map}} = \argmin_{\beta} \sum_{i=1}^N ( \hat{Y}'_i - \beta^\tran \hat{X}'_i)^2.
\end{align*}
The standard error of the map-only coefficient estimate will be small if $N$ is large, resulting in a narrow map-only 95\% confidence interval. However, we employ the $\gamma$ notation to emphasize that the map-only estimator may be \textit{biased}; if the remotely sensed values $\hat{X}'_i$ or $\hat{Y}'_i$ are inaccurate, $\hat{\gamma}_{\text{map}}$ might be biased in the sense that as the sample size goes to infinity, $\hat{\gamma}_{\text{map}}$ will converge (in probability) to some $\gamma_* \neq \beta_*$. Thus, the map-only confidence interval is not guaranteed to have the appropriate coverage rate (in this case, 95\%) of the true parameter of interest. This is because the map-only estimator does not account for error or uncertainty in the model used to generate the map product predictions. We note that the map-only estimator is analogous to the pixel-counting area estimator described in Section \ref{sec:introduction}, but the map-only estimator applies to regression coefficient estimation rather than area estimation.

\paragraph{Prediction-Powered Inference: Map + ground truth.} Introduced in 2023, \textbf{Prediction-Powered Inference (PPI)} \cite{angelopoulos2023prediction} is a method that computes point estimates and confidence intervals for parameters of interest using a small calibration set of ground truth data points and a much larger set of machine learning-generated predictions. The main idea is that PPI assesses the bias in the map product using the ground truth data and then corrects it. For regression coefficient estimation, we use variants of PPI that were developed in \citet{Kluger25GeneralizingPPI} and have origins in \citet{chen2000unified,zrnic2024note}. (PPI can also be applied to mean and area estimation; see Appendices \ref{appendix:ppi-mean-estimation} and \ref{appendix:area-estimation}.)

PPI uses the calibration data to correct for bias in the map-based estimator. First, it computes the regression coefficient estimate using the remote sensing predictions at the $n$ calibration data points:
\begin{align*}
    \hat{\gamma}_{\text{calib}} = \argmin_{\beta} \sum_{i=1}^n ( \hat{Y}_i - \beta^\tran \hat{X}_i)^2.
\end{align*}
Note that $\hat{\gamma}_{\text{calib}}$ has the same bias as the map-only estimator $\hat{\gamma}_{\text{map}}$, so we can use $\hat{\gamma}_{\text{calib}}$ in conjunction with the GT-only estimator $\hat{\beta}_{\text{calib}}$ to estimate the bias of the map-only estimator. 

The PPI regression coefficient estimator corrects the map-only estimator $\hat{\gamma}_{\text{map}}$ for bias by using the difference between the ground truth and map estimates on the calibration set $(\hat{\beta}_{\text{calib}} - \hat{\gamma}_{\text{calib}})$. The final PPI estimator is 
\begin{align*}
    \hat{\beta}_{\text{PPI}} = \underbrace{\hat{\gamma}_{\text{map}}}_{\text{map-only estimate}} + \underbrace{(\hat{\beta}_{\text{calib}} - \hat{\gamma}_{\text{calib}})}_{\text{bias correction}}.
\end{align*}

The PPI estimator can be improved by computing a tuning matrix $\hat{\Omega}$ that minimizes the variance of each component of the estimated coefficient vector. The value of $\hat{\Omega}$ depends on the quality of the map product relative to the ground truth. The tuned PPI regression coefficient estimator is
\begin{align}\label{ppi-regression-equation}
    \hat{\beta}_{\text{PPI}} = \underbrace{\hat{\Omega} \hat{\gamma}_{\text{map}}}_{\text{map-only estimate}} + \underbrace{(\hat{\beta}_{\text{calib}} - \hat{\Omega} \hat{\gamma}_{\text{calib}})}_{\text{bias correction}}.
\end{align}
The tuned $\hat{\beta}_{\text{PPI}}$ is still an unbiased estimator for the true coefficient $\beta_{*}$. The analytic formula for an optimal choice of tuning matrix is presented in Appendix \ref{appendix:FormulaForPPIoptTuningAndCov} for interested readers. We remark that rearranging terms gives $ \hat{\beta}_{\text{PPI}} = \hat{\beta}_{\text{calib}} + \hat{\Omega} (\hat{\gamma}_{\text{map}}-\hat{\gamma}_{\text{calib}})$, and thus the PPI estimator has notable parallels to the regression mean estimator (Appendix \ref{appendix:regression-estimator}).

We compute a PPI confidence interval for the regression coefficients using the \textbf{prediction-powered percentile bootstrap} method~\cite{zrnic2024note, Kluger25GeneralizingPPI}. In particular, we resample both the calibration set and map data points with replacement $B$ times, and for each $1 \leq i \leq B$, we compute the point estimate $\hat{\beta}_{\text{PPI}}^{(i)}$ using the $i$th resampled dataset. For each $j=1,\dots,p$, the $(1-\alpha)$-level confidence interval for $\beta_j$ has endpoints that are given by the $\alpha/2$ percentile and $1-\alpha/2$ percentile of $j$th entries of the $B$ estimated vectors $\hat{\beta}_{\text{PPI}}^{(1)}, \hat{\beta}_{\text{PPI}}^{(2)}, \dots, \hat{\beta}_{\text{PPI}}^{(B)}$. Larger values of $B$ result in more algorithmically stable confidence intervals, at the cost of increased computational time. Note that an alternative approach to constructing confidence intervals for the estimator in Equation \eqref{ppi-regression-equation} would be to derive an asymptotic normal approximation and variance formula for $\hat{\beta}_{\text{PPI}}$ (e.g., such an approach was developed in \citet{chen2000unified}; we derive it in Appendix \ref{appendix:FormulaForPPIoptTuningAndCov}). However, in this paper we focus on bootstrap-based approaches to constructing confidence intervals because they generalize more readily to nonstandard regression models and can be modified to apply to settings where the calibration set is selected using unequal probability sampling, stratified random sampling, or cluster random sampling \cite{Kluger25GeneralizingPPI}.

\subsection{Logistic regression} When $Y$ is a binary variable, we can similarly use PPI to estimate the logistic regression coefficients $\beta$ that give the best fit to the model
\begin{align*}
    \mathbb{P}(Y=1|X) = \frac{1}{1+e^{-\beta^\tran X}}.
\end{align*} 
(We assume that the first element of the vector $X$ is a constant 1, so that the first regression coefficient $\beta_0$ is an intercept term.)

The ground truth-only estimator using the $n$ ground truth calibration data points is
\begin{align*}
    \hat{\beta}_{\text{calib}} = \argmin_{\beta} \Bigl\{ \sum_{i=1}^n \log(1+e^{ \beta^\tran X_i})-Y_i \beta^\tran X_i \Bigr\}.  
\end{align*}

The map-only estimator using the $N$ map data points is
\begin{align*}
    \hat{\gamma}_{\text{map}} = \argmin_{\beta} \Bigl\{ \sum_{i=1}^N \log(1+e^{ \beta^\tran \hat{X}'_i})-\hat{Y}'_i \beta^\tran \hat{X}'_i \Bigr\}. \end{align*} 

The estimator using the remote sensing predictions at the $n$ calibration data points is
\begin{align*}
    \hat{\gamma}_{\text{calib}} = \argmin_{\beta} \Bigl\{ \sum_{i=1}^n \log(1+e^{ \beta^\tran \hat{X}_i})-\hat{Y}_i \beta^\tran \hat{X}_i \Bigr\}.
\end{align*}

Similar to the linear regression case, the PPI logistic regression coefficient estimate is obtained by plugging these values into Equation \eqref{ppi-regression-equation}, and the confidence interval is obtained using the percentile bootstrap. In the case of logistic regression, analytic formulas for an optimal choice of tuning matrix and an estimator of the covariance matrix of $\hat{\beta}_{\text{PPI}}$ are also presented in Appendix \ref{appendix:FormulaForPPIoptTuningAndCov}.

\subsection{Error-in-$Y$ vs. Error-in-$X$ vs. Error-in-both}\label{sec:error-regimes}

PPI for regression coefficient estimation can be applied even when only a subset of the variables are remotely sensed, while the others have ground truth labels at all $N$ locations. We study PPI in three different regression regimes of interest, using the same nomenclature and categorization as \citet{proctor2023parameter}. In the first regime, which we call \textbf{error-in-$Y$}, the investigator wishes to leverage a map product for the response variable, while ground truth data for the covariates are widely available. In the second regime, which we call \textbf{error-in-$X$}, the investigator wishes to leverage a map product for some of the covariates, while ground truth data for the response variable is widely available. In the third regime, which we call \textbf{error-in-both}, ground truth labels are sparsely available for both the response variable and some of the covariates, and the investigator wishes to leverage a map product for these sparsely measured variables.

We consider these three regimes separately for two reasons. First, many existing statistical methods either only apply to the error-in-$X$ regime or only apply to the error-in-$Y$ regime. For example, some earlier variants of PPI \cite{angelopoulos2023prediction,angelopoulos2023ppi++} are only designed for the error-in-$Y$ regime, while many of the traditional methods in the measurement error literature \cite{CarrolStefanski06} are only designed for error-in-$X$ regime. The PPI algorithm in \citet{angelopoulos2023ppi++} debiases an empirical loss computed from map predictions; it requires the debiased empirical loss to be convex, which can be violated when there is error in $X$. By contrast, the variant of PPI presented in this paper \cite{Kluger25GeneralizingPPI} debiases the map-only regression coefficient estimator directly. We emphasize that as a result, this PPI algorithm can be applied to all three regimes. 

Second, the bias of the coefficient is thought to have different behavior and severity in the three regimes. The map-only estimator bias is hypothesized to be least severe in the error-in-$Y$ regime and most severe in the error-in-both regime. For example, if the map error is assumed to be additive noise, the map-only linear regression estimator will have no bias in the error-in-$Y$ regime \cite{CarrolStefanski06} but in the error-in-$X$ regime it will often be biased towards zero (attenuation bias), and can sometimes be biased away from zero \cite{MythsEpi5Paper,Kluger24BiasDirection}. In the error-in-both regime, the bias issues from the error-in-$Y$ and error-in-$X$ regimes can aggregate.


We emphasize that in all three regimes, we are interested in identifying the relationship between response variable $Y$ and covariates $X$. Thus, both $Y$ and $X$ contain variables of interest; unlike in the regression mean estimator described in Appendix \ref{appendix:regression-estimator}, $X$ is not merely a set of auxiliary variables extracted from satellite imagery to be used to help estimate $Y$.

\paragraph{Examples} All three of these regression settings are commonly seen in the remote sensing literature. We list a few examples here. In each example, we consider only the remotely sensed variable(s) to have error. For error-in-$Y$, \citet{taylor2023fertilizer} regressed remotely sensed algal blooms against fertilizer use data from the US Geological Survey. For error-in-$X$, studies have regressed crop yield data from the National Bureau of Statistics of China against remotely sensed air pollution \cite{he2022marked}, and US flood insurance claims from FEMA's National Flood Insurance Program Redacted Claims Dataset against remotely sensed wetland loss \cite{taylor2022wetlands}. For error-in-both, studies have regressed remotely sensed maize and soybean yields against remotely sensed cover crop adoption in the US Corn Belt \cite{deines2023recent}, remotely sensed maize and soybean yields against remotely sensed conservation tillage in the US Corn Belt \cite{deines2019satellites}, and remotely sensed crop greenness against remotely sensed nitrogen oxide levels in several agricultural regions \cite{lobell2022globally}.


Throughout this paper, we consider data to be ``ground truth" if they are reliable observations of a variable of interest; ground truth data may be either sparse or widely available. In many applications, such as the examples listed above, remotely sensed data are error-prone proxies for a variable of interest that is sparsely observed (e.g., remotely sensed air pollution is a proxy for true air pollution, which is often only monitored in a small number of locations). However, in some use cases, a remotely sensed variable may itself be a variable of interest. For example, we may wish to understand the relationship between the remotely sensed Normalized Difference Vegetation Index (NDVI) and some response variable, such as crop yield \cite{groten1993ndvi, balaghi2008empirical, mkhabela2011crop}. In such cases, we would treat the remotely sensed data as ground truth that is widely available, and apply PPI accordingly.

\subsection{Evaluating performance using effective sample size}

We use \textbf{effective sample size} to compare the effectiveness of regression coefficient estimation methods. Suppose a GT-only confidence interval of width $w$ is generated using $n$ ground truth data points. If another estimation method (such as PPI) uses $n$ ground truth data points (along with a map product) to generate a confidence interval of width $w'$, then the effective sample size of the new method is 
\begin{align*}
    n_{\text{effective}} = n \cdot \left(\frac{w}{w'}\right)^2.
\end{align*}
This is the number of ground truth data points that would be required to generate a GT-only confidence interval of width $w'$.

\section{Examples of regression coefficient estimation}\label{sec:examples}

In this section, we apply PPI to estimate regression coefficients in four use cases associating deforestation, tree cover, housing price, and forest cover to other geospatial covariates. We compare point estimates and confidence intervals for the map-only estimator, GT-only estimator, and PPI estimator. Note that our regressions are not meant to contain a comprehensive list of covariates; our examples are primarily a means to illustrate the strengths and weaknesses of each method rather than meant to be interpreted scientifically.

PPI can be applied when the response variable, the covariates, or both are remotely sensed, as long as each remotely sensed variable has a small amount of ground truth data for calibration. In the deforestation and tree cover regressions, the response variable is remotely sensed (error in $Y$), but ground truth data are widely available for all covariates. In the housing price regression, the response variable has ground truth widely available, but two of the covariates are remotely sensed (error in $X$). In the forest cover regression, both the response variable and one of the covariates are remotely sensed (error in both $X$ and $Y$).

\begin{figure}[p]
\centering
\includegraphics[width=\textwidth]{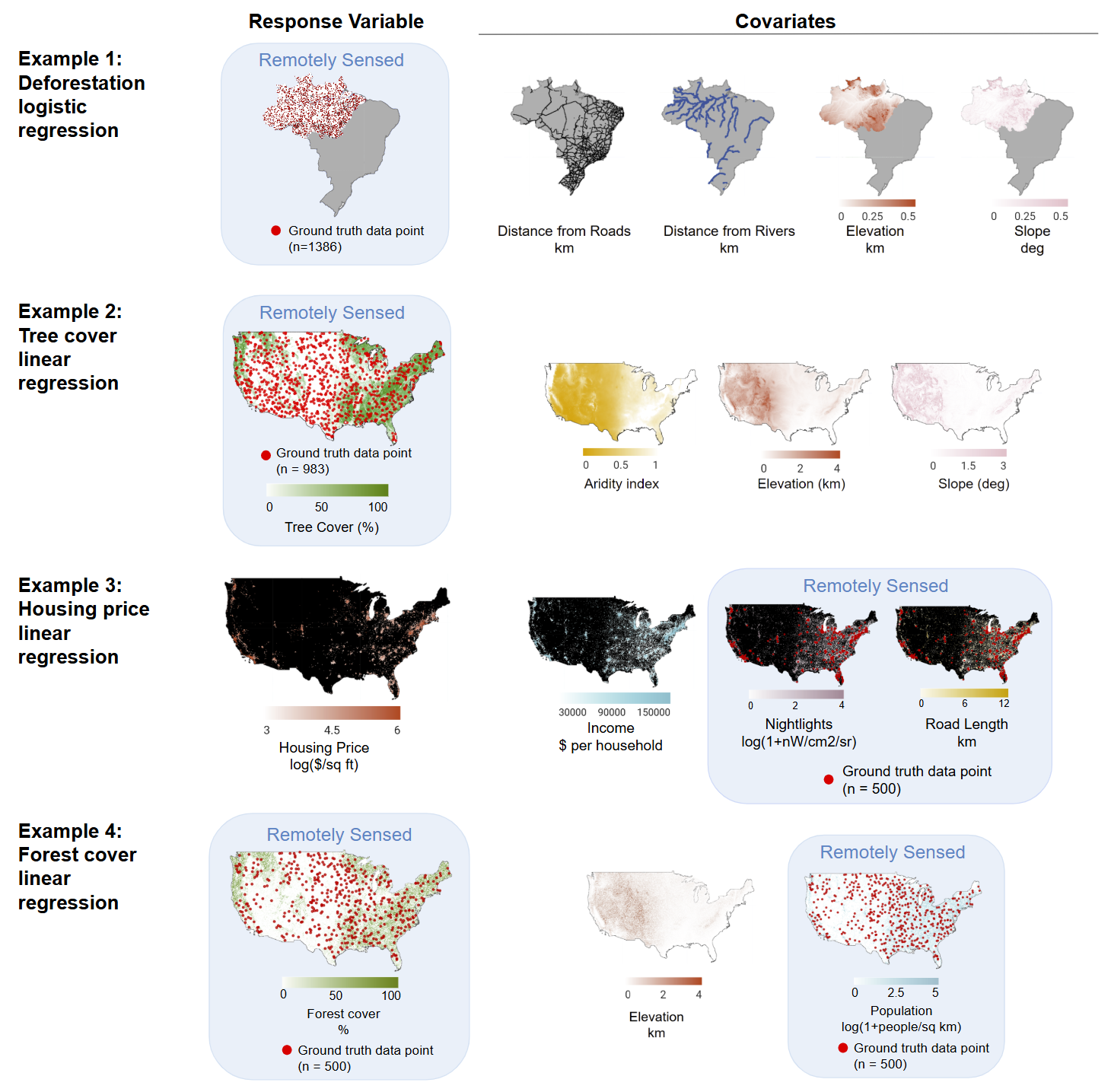}
\caption{\textbf{Response variable and covariates for the four use cases.} We consider use cases with remotely sensed response variable (Examples 1 and 2), remotely sensed covariates (Example 3), or both (Example 4).}\label{dataset-maps}
\end{figure}

\subsection{Experimental setup}\label{subsection:use-case-datasets}
We first summarize the regression specifications and datasets for the four use cases. Maps for each use case are shown in Figure \ref{dataset-maps}. For more detailed descriptions of the datasets, see Appendix \ref{appendix:datasets}. 

\paragraph{Example 1: Deforestation logistic regression} We estimate the logistic regression coefficients for the association between Brazilian Amazon deforestation in 2000--2015 and several covariates: distance from major roads, distance from major rivers, elevation, and slope. We use logistic regression because the response variable (deforested or non-deforested) is binary.  
Given $n$ data points, the logistic regression model is 
\begin{align*}
    \log\frac{p}{1-p} = \beta_0 + \beta_{\text{distRoad}}\textit{X}_{\text{distRoad}} + \beta_{\text{distRiver}}\textit{X}_{\text{distRiver}} + \beta_{\text{elevation}}\textit{X}_{\text{elevation}} +
    \beta_{\text{slope}}\textit{X}_{\text{slope}}
\end{align*}
where $p$ is a $n \times 1$ vector of probabilities, $\textit{X}_j$ is a $n \times 1$ vector of covariate $j$, and $\beta_j$ is the scalar logistic regression coefficient for covariate $j$. We wish to estimate the $\beta_j$ giving the best fit to the above model. A one-unit increase in covariate $j$ corresponds to a $\beta_j$ change in the log-odds of the response variable. The scalar $\beta_0$ is the intercept, corresponding to the log-odds of the response when the covariates $\textit{X}_j$ are zero.

The deforestation response variable is remotely sensed while the covariates are treated as ground truth. Each data point is a pixel with 30 m resolution. For remotely sensed deforestation, we create a binary deforestation map product using $N=963548$ data points selected via simple random sampling from the NASA Global Forest Cover Change (GFCC) map \cite{townshend2016global} in the study area in 2000 and 2015. For ground truth deforestation, we use a simple random sample of $n=1386$ data points from an Amazon deforestation dataset created by \citet{bullock2020satellite} using time-series analysis of Landsat imagery and high-resolution imagery from Google Earth. For the covariates, we use a map of Brazilian federal roads in 2000 \cite{de2021transport}, the WWF HydroSHEDS Free Flowing Rivers Network map \cite{grill2019mapping}, and the NASA Digital Elevation Model \cite{jpl2020}. See Appendix \ref{appendix:datasets-deforestation} for more information about the data. 

\paragraph{Example 2: Tree cover linear regression} We estimate linear regression coefficients for the association between 2021 tree cover percentage in the contiguous United States and several covariates: aridity index, elevation, and slope. We use linear regression rather than logistic regression because the response variable (tree cover percentage) is continuous rather than binary. Given $n$ data points, the linear regression model is 
\begin{align*}
    Y = \beta_0 + \beta_{\text{aridity}}\textit{X}_{\text{aridity}} + \beta_{\text{elevation}}\textit{X}_{\text{elevation}} +
    \beta_{\text{slope}}\textit{X}_{\text{slope}} + \varepsilon
\end{align*}
where $Y$ is a $n \times 1$ vector of the response variable, $\textit{X}_j$ is a $n \times 1$ vector of covariate $j$, and $\beta_j$ is the scalar linear regression coefficient for covariate $j$. We wish to estimate the $\beta_j$ that give the best fit to the above model. A one-unit increase in covariate $j$ corresponds to a $\beta_j$ change in the response variable. The scalar $\beta_0$ is the intercept, corresponding to the response value when the covariates $\textit{X}_j$ are zero. 

The tree cover response variable is remotely sensed while the covariates are treated as ground truth. Each data point is a pixel with 30 m resolution. For remotely sensed tree cover, we use $N=983238$ map product data points selected via simple random sampling from the 2021 USFS Tree Canopy Cover (TCC) product \cite{forest2023usfs}. For ground truth tree cover, we manually label $n=983$ data points by downloading high-resolution satellite images from Google Maps (at locations selected via simple random sampling) and visually estimating the proportion of each 30 m by 30 m image that is covered by trees. For the covariates, we use a map of the average Global Aridity Index from 1970-2000 \cite{trabucco2019global} and the NASA Digital Elevation Model \cite{jpl2020}. See Appendix \ref{appendix:datasets-tree-cover} for more information about the data.

\paragraph{Example 3: Housing price linear regression} We estimate linear regression coefficients for the association between housing price in the United States and household income, nighttime light intensity, and road length. The linear regression model is 
\begin{align*}
    Y = \beta_0 + \beta_{\text{income}}\textit{X}_{\text{income}} + \beta_{\text{nightlight}}\textit{X}_{\text{nightlight}} +
    \beta_{\text{roadLen}}\textit{X}_{\text{roadLen}} + \varepsilon
\end{align*} 
using similar notation as the previous linear regression example. 

Two of the covariates (nightlights and road length) are remotely sensed, but ground truth data are widely available for the response variable (housing price) and one covariate (income). For all variables, we use the MOSAIKS dataset \cite{rolf2021generalizable} which contains an unequal probability sample of data points in the United States. (Specifically, the data points are sampled using an unequal probability sampling design, such that data points at locations with higher population density are more likely to be selected.) Each data point is a grid cell with roughly 1 km resolution. There are $N=46418$ data points that have both ground truth and remotely sensed values for housing price (log dollars per square foot), household income (dollars), nighttime light intensity ($\log(1 + \text{nW}/\text{cm}^2/\text{sr})$), and road length (km). See Appendix \ref{appendix:datasets-housing-price} for more information about the data.

For housing price and income, we use the ground truth values for all $46418$ data points. For nightlights and road length, we take a simple random subsample of $n=500$ data points as the ground truth calibration set (and use remotely sensed values for the remaining data points). For the GT-only method, we use only the $n=500$ ground truth data points for all four variables. We also compare the regression coefficient point estimates from our methods against the ``true coefficients" computed by running linear regression on the full set of $46418$ ground truth data points for all four variables. 

\paragraph{Example 4: Forest cover linear regression}
We estimate linear regression coefficients for the association between forest cover in the United States and two covariates: elevation and population. The linear regression model is 
\begin{align*}
    Y = \beta_0 + \beta_{\text{elevation}}\textit{X}_{\text{elevation}} + \beta_{\text{population}}\textit{X}_{\text{population}} + \varepsilon
\end{align*}
using similar notation as the previous linear regression examples.

Both the response variable (forest cover) and one of the covariates (population) are remotely sensed, while ground truth data are widely available for the other covariate (elevation). Similar to the previous example, we use the MOSAIKS dataset \cite{rolf2021generalizable}. The dataset also provides a simple random sample of data points in the United States, where each data point is a grid cell with roughly 1 km resolution. After removing data points with missing variables, there are $N=67968$ data points that have both ground truth and remotely sensed values for forest cover percentage, elevation (km), and population ($\log(1+\text{people})/\text{km}^2$). See Appendix \ref{appendix:datasets-forest-cover} for more information about the data.

For elevation, we use the ground truth values for all $67968$ data points. For forest cover and population, we take a simple random subsample of $n=500$ data points as the ground truth calibration set (and use remotely sensed values for the remaining data points). For the GT-only method, we use only the $n=500$ ground truth data points for all three variables. We also compare the regression coefficient point estimates from our methods against the ``true coefficients" computed by running linear regression on the full set of $67968$ ground truth data points for all four variables. 

\subsection{Results}

\begin{figure}[]
\centering
\includegraphics[width=0.9\textwidth]{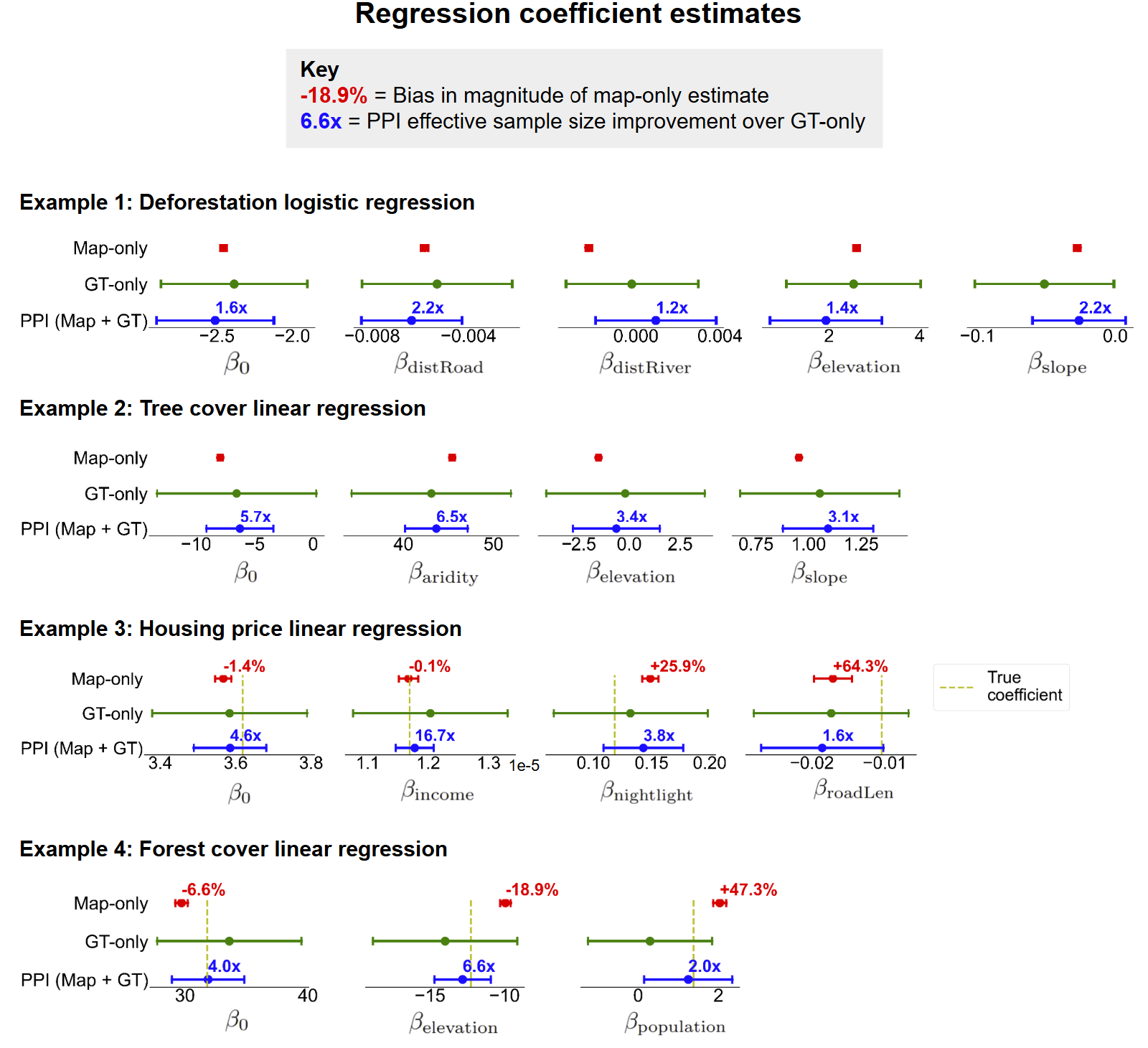}
\caption{\textbf{Regression coefficient point estimates and 95\% confidence intervals for the four use cases.} PPI results in narrower 95\% confidence intervals than using ground truth only. PPI effective sample size improvements (relative to GT-only) are shown in blue. For the two examples where the true coefficient is known, the bias in magnitude of each map-only estimate is shown in red.}\label{regression-results}
\end{figure}

The regression coefficient point estimates and 95\% confidence intervals for all four use cases are shown in Figure \ref{regression-results}. We compare results for the map-only, GT-only, and PPI estimators. (For all experiments in this section, we use $B=2000$ bootstrap iterations to construct PPI confidence intervals.)

\paragraph{Map-only estimator is biased and its confidence intervals are too narrow.} For all coefficients, the map-only confidence intervals are much narrower than those for GT-only and PPI. Unlike the GT-only and PPI intervals, the map-only intervals are not guaranteed with 95\% probability to contain the true coefficients (i.e., the GT-only coefficients we would compute if we had access to ground truth data at all $N$ map product data point locations).

For instance, in Example 1 (deforestation logistic regression), the map-only estimate for the distance from river coefficient has a very narrow confidence interval around $-0.002$ and falls outside the PPI interval. Meanwhile, the GT-only and PPI confidence intervals are much wider, capturing the true uncertainty in the data, and both intersect with zero. Using only the maps would lead us to conclude that there is a significantly negative association between deforestation and distance from the nearest river, while the two methods that use ground truth data points suggest that there may be no association. 

In Examples 3 and 4 (housing price and forest cover linear regressions), we have access to the true coefficients computed from the MOSAIKS dataset. We find that for all but one coefficient, the map-only confidence interval does not contain the true coefficient. 
We compute the bias (in percentages) of each map-only coefficient point estimate by subtracting the true coefficient from the map-only coefficient, then dividing the result by the true coefficient. Negative values of bias indicate the map-only coefficient is biased toward zero, and positive values indicate the map-only coefficient is biased away from zero. In our examples, the biases of the map-only coefficients range from $-18.9\%$ to $+64.3\%$. 

The single map-only confidence interval that contains the true coefficient is for income in the housing price linear regression. While there are no guarantees, a map-only estimator could happen to be unbiased for a variety of reasons. In the case of an error-in-$X$ linear regression, the coefficient estimator corresponding to a covariate for which only ground truth observations are used will be unbiased if that covariate is uncorrelated with both the map predictions and the map prediction errors of the other covariates (see Formula (5) in \citet{Kluger24BiasDirection}). We suspect the map-only estimate for the income coefficient has low bias in our experiment because (i) ground truth income data are used for all samples, (ii) income is uncorrelated with the errors in the nightlights and road length predictions, and (iii) income is only weakly correlated with the nightlights and road length predictions themselves. 

\paragraph{PPI uses ground truth data to correct for map bias.} The PPI estimator is unbiased, and the PPI confidence intervals have theoretical guarantees of containing the true coefficients with 95\% probability. In Examples 3 and 4, the GT-only and PPI confidence intervals contain all of the true coefficients. PPI successfully uses the ground truth calibration set to correct for bias from the map product and account for uncertainty due to map product errors. 

\paragraph{PPI confidence intervals outperform the GT-only method.} PPI uses the map product along with the ground truth to reduce uncertainty in the coefficient estimates, resulting in significantly narrower confidence intervals and larger effective sample sizes compared to the GT-only method. In our experiments, the PPI effective sample sizes range from $1.2\times$ to $16.7\times$ the ground truth calibration set size $n$. 

The largest PPI effective sample size improvement ($16.7\times$) is for the income coefficient in Example 3. We suspect that this is because the map-only income coefficient was essentially unbiased, so the PPI estimator for this coefficient involved a very small bias correction term, and therefore was similar to the low variance map-only estimator. 
The next largest PPI effective sample size improvements are for the elevation coefficient ($6.6\times$) in Example 4 and the aridity coefficient ($6.5 \times$) and intercept ($5.7 \times$) in Example 2 (tree cover linear regression). In contrast, adding map data did not reduce the confidence interval size significantly for the distance to river ($1.2\times$) and elevation ($1.4\times$) coefficients in Example 1.

\section{Simulating the role of map error}\label{sec:simulations}

To understand the effect of map product quality on PPI regression coefficient estimates, we simulate adding noise and bias to the map products in Example 4 (forest cover linear regression) from the previous section. We compare the resulting point estimates and 95\% coefficient confidence intervals at different levels of map noise and bias using map-only, GT-only, and PPI estimators.

\subsection{Experimental Setup}

Similar to Example 4, we use the MOSAIKS dataset to regress forest cover against elevation and population, where forest cover and population are remotely sensed. We use the same simple random subsample of $n=500$ ground truth data points as in the original experiment. However, instead of using the MOSAIKS map product datasets for forest cover and population, we take the full MOSAIKS ground truth dataset (67968 data points) and use it to create simulated ``map products" with different levels of noise and bias. We experiment with adding noise or bias to the population map (error-in-$X$), the forest cover map (error-in-$Y$), or both (error-in-both). No error is added to the elevation dataset. 

\paragraph{Simulating Noise} For the ground truth population covariate $X_{\text{pop}}$, we define ``noise at level $c$" as additive Gaussian noise with mean zero and standard deviation equal to $c$ times $\sigma_{X_{\text{pop}}}$, the standard deviation of all $X_{\text{pop}}$ values in the full ground truth dataset. The resulting simulated map product population value at each data point is 
\begin{align*}
    X_{\text{pop}}^{\text{map}} = X_{\text{pop}} + c \cdot \mathcal{N}(0, \sigma^{2}_{X_{\text{pop}}}).
\end{align*}
Similarly, for the forest cover response variable $Y$, the simulated map product value at each data point is
\begin{align*}
    Y^{\text{map}} = Y + c \cdot \mathcal{N}(0, \sigma^{2}_{Y}).
\end{align*}
We experiment with noise levels $c$ from 0 to 1 at increments of 0.1. When $c=0$, the simulated map is the same as the full ground truth map. 

\paragraph{Simulating Bias} To simulate biased predictions for the population covariate, we fit a square root regression relating the ground truth values $X_{\text{pop}}$ to the corresponding MOSAIKS map product values $X_{\text{pop}}'$, using all $N=67968$ data points. We obtain the regression $\hat{X}_{\text{pop}}' = 1.74\sqrt{X_{\text{pop}}} - .22$. We can treat $\hat{X}_{\text{pop}}'$ as a biased prediction for $X_{\text{pop}}$. We choose a square root curve as an example to simulate nonlinear bias in map products, as we often observe machine learning predictions to saturate at high values of a variable. (Another model of bias, linear mean reversion, is explored through simulations in Appendix \ref{appendix:mean-reversion-bias}.)

Then we create a simulated map product with ``bias level $c$" by interpolating between $X_{\text{pop}}$ and $\hat{X}_{\text{pop}}'$ to get
\begin{align*}
    X_{\text{pop}}^{\text{map}} = c\hat{X}_{\text{pop}}' + (1-c)X_{\text{pop}}.
\end{align*} Similarly, for the forest cover response variable $Y$, we use the MOSAIKS map product $Y'$ to fit the square root regression $\hat{Y}' = 7.90\sqrt{Y} - 3.20$. The simulated map product value at each data point is 
\begin{align*}
    Y^{\text{map}} = c\hat{Y}' + (1-c)Y.
\end{align*} We experiment with bias levels $c$ from 0 to 1 at increments of 0.1. When $c=0$, the simulated map is the same as the full ground truth map. When $c=1$, the simulated map is the same as the biased prediction $\hat{X}_{\text{pop}}'$ or $\hat{Y}'$.

\begin{figure}[]
\centering
\includegraphics[width=0.9\textwidth]{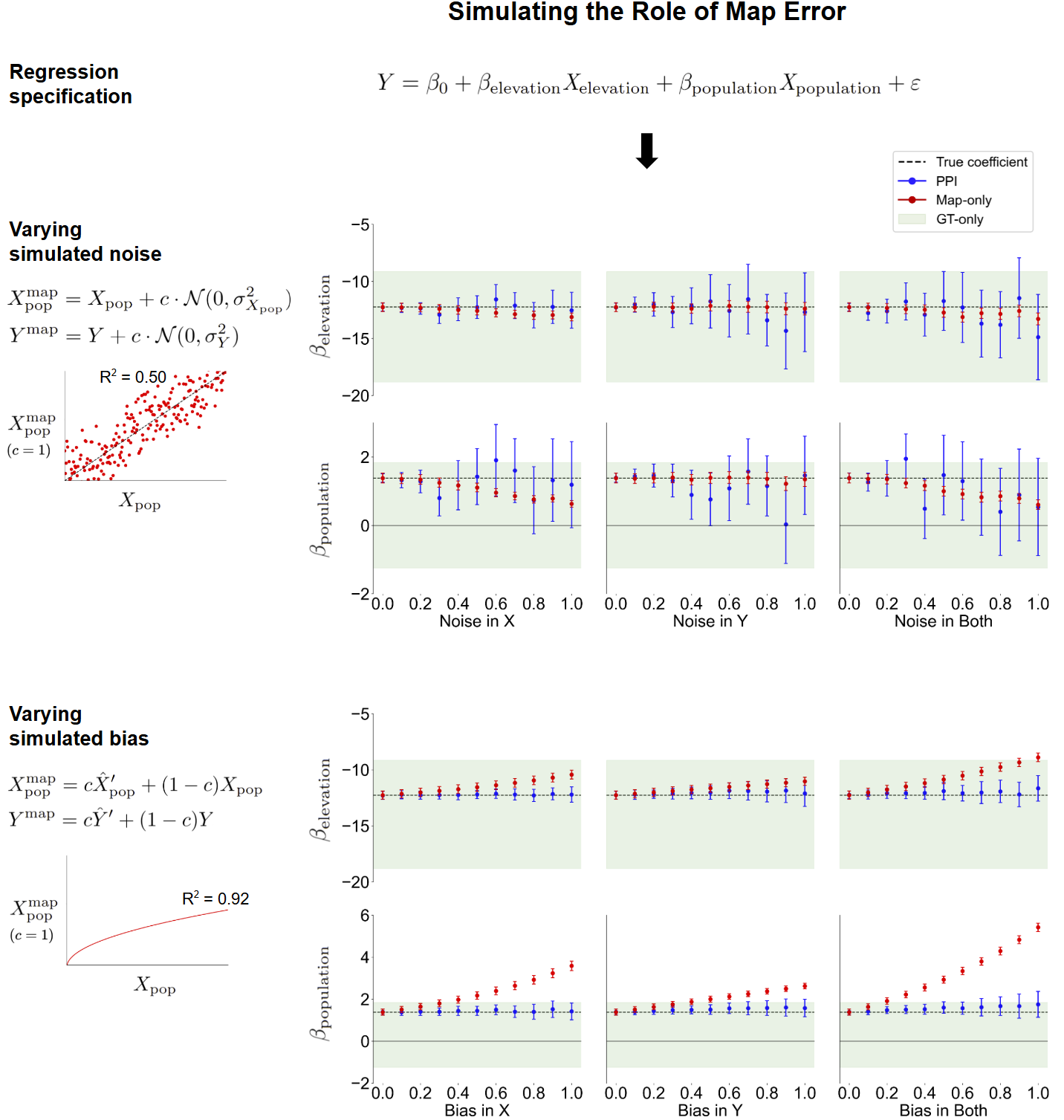}
\caption{\textbf{Simulation estimating forest cover linear regression coefficient 95\% confidence intervals under different levels of map noise and map bias.} In the bias simulations, $\hat{X}_{\text{pop}}'$ and $\hat{Y}'$ are nonlinearly biased proxies for true $X_{\text{pop}}$ and $Y$. Even with map noise or bias, PPI is unbiased and produces narrower confidence intervals than the GT-only method. Map-only estimates become increasingly biased as noise in the simulated map products for $X_{\text{pop}}$ increases or bias in the simulated map products for $X_{\text{pop}}$ or $Y$ increases. We also display the $R^2$ coefficient of determination for $X_{\text{pop}}^{\text{map}}$ at $c=1$ noise and bias levels.}\label{simulations}
\end{figure}

\subsection{Results}
The coefficient point estimates and 95\% confidence intervals at each noise and bias level are shown in Figure \ref{simulations}. We compare results for the map-only, GT-only, and PPI estimators. (For all experiments in this section, we use $B=200$ bootstrap iterations to construct PPI confidence intervals.)

\paragraph{GT-only confidence interval is wide.} First, the GT-only point estimate and confidence interval stay constant regardless of noise and bias, because they use a constant set of ground truth data points. The confidence interval contains the true coefficients for both covariates ($-12$ for elevation and 1.4 for population). However, due to the small number of ground truth data points ($n=500$), the GT-only interval for $\beta_{\text{pop}}$ also contains zero, so it does not detect a statistically significant association between forest cover and population. The GT-only interval is wider than the PPI and map-only intervals; exactly how much wider depends on the noise and bias levels. 

\paragraph{Both noise and bias lead to biased map-only estimates.} Second, the map-only estimates suffer from bias when either noise or bias is added to the variables. In the noise experiments, the map-only estimates for $\beta_{\text{pop}}$ and $\beta_{\text{elevation}}$ become biased when there is noise in $X_{\text{pop}}$. As noise increases, the map-only estimates for $\beta_{\text{pop}}$ decrease in magnitude and become biased toward zero, because additive Gaussian noise in a covariate results in attenuation bias in the corresponding regression coefficient (see Section \ref{sec:error-regimes}). The map-only estimates for $\beta_{\text{elevation}}$ conversely become larger in magnitude as noise in $X_{\text{pop}}$ increases. 

Furthermore, the map-only confidence intervals are narrow because the number of map product data points is large ($N = 67968$), but unlike the PPI intervals, they fail to capture the true estimation uncertainties due to map imperfections. The confidence intervals for $\beta_{\text{pop}}$ do not contain the true coefficient for $X_{\text{pop}}$ noise levels $c \geq 0.4$ (corresponding to $R^2 \leq 0.92$ for the simulated population map). The confidence intervals for $\beta_{\text{elevation}}$ do not contain the true coefficient for $X_{\text{pop}}$ noise levels $c \geq 0.6$ ($R^2 \leq 0.82$). By $X_{\text{pop}}$ noise levels of $c = 0.8$ ($R^2=0.68$), the map-only estimate of $\beta_{\text{pop}}$ is roughly half of the true coefficient. In contrast, noise in $Y$ (forest cover) does not result in biased map-only estimates, because zero-centered noise in a response variable does not bias regression coefficients.

Whether bias is in $X_{\text{pop}}$, $Y$, or both, the map-only estimates become biased. Similar to in the noise experiments, when we add bias the map-only confidence intervals remain narrow but fail to capture the actual estimation uncertainty. The map-only confidence intervals do not contain the true coefficient for $\beta_{\text{pop}}$ at bias levels $c \geq 0.2$ ($R^2 \leq 0.997$ for simulated population and forest cover maps) or the true coefficient for $\beta_{\text{elevation}}$ at bias levels $c \geq 0.4$ ($R^2 \leq 0.987$ for $X_{\text{pop}}$ and $R^2 \leq 0.988$ for $Y$). When the bias level of $X_{\text{pop}}$ is at $c=1$ (in other words, $X_{\text{pop}}^{\text{map}} = \hat{X}_{\text{pop}}'$)---corresponding to $R^2 = 0.92$ of the map compared to ground truth---treating the map as if it were 100\% accurate would nearly double the estimate for $\beta_{\text{pop}}$. The direction of bias in coefficient estimates varies depending on the type of bias in the map and can be difficult to predict in general. In our bias experiments, the map-only estimates for $\beta_{\text{pop}}$ become biased away from zero while those for $\beta_{\text{elevation}}$ become biased toward zero.

We reported the $R^2$ of map products above because $R^2$ is a common evaluation metric for remote sensing maps (MSE is another common metric, but its values are variable-specific). We draw attention to the fact that map $R^2$ can be high ($> 0.9$)---in other words, the map appears very accurate---but the map-only estimator still yields significantly biased coefficients.

\paragraph{Performance of PPI.} Finally, our simulations allow us to see clearly the advantages and limitations of PPI. At almost all noise and bias levels across experiments, the PPI confidence intervals contain the true coefficients for both covariates. (The only exceptions are noise levels $c=0.3$ in $X_{\text{pop}}$ ($R^2=0.96$), $c=0.9$ in $Y$ ($R^2=0.19$), and $c=0.4$ in both ($R^2=0.92$ for $X_{\text{pop}}$ and $R^2=0.84$ for $Y$).) This is expected, as PPI uses the small set of ground truth data points to correct for error and quantify uncertainty in the map product, and the PPI confidence intervals are guaranteed to contain the true coefficient with 95\% probability. 

In contrast to using GT-only, PPI estimates $\beta_{\text{pop}}$ and $\beta_{\text{elevation}}$ much more precisely, allowing us to correctly conclude that there is a positive association between forest cover and population. The PPI estimate for $\beta_{\text{pop}}$ is only indistinguishable from zero at noise levels of $c \ge 0.8$ in $X_{\text{pop}}$ ($R^2 \leq 0.68$), $c \ge 0.5$ in $Y$ ($R^2 \leq 0.75$), or $c \ge 0.4$ in both ($R^2 \leq 0.92$ for $X_{\text{pop}}$ and $R^2 \leq 0.84$ for $Y$). The PPI estimates are statistically significant at all bias levels in our experiment.

The effectiveness of PPI is dependent on the quality of the map product. When the noise or bias level is low, the PPI intervals are significantly narrower than the GT-only interval, because the simulated map product is close to ground truth and provides useful information for estimating the regression coefficients. As the noise level increases and the quality of the map deteriorates, the PPI intervals grow wider, approaching the GT-only interval. For example, when the noise level is $c=0.2$ in both $X_{\text{pop}}$ and $Y$ ($R^2=0.98$ for $X_{\text{pop}}$ and $R^2=0.96$ for $Y$), the PPI effective sample size outperforms the GT-only method by a factor of $21.6\times$ for elevation and $8.8\times$ for population. When the noise level increases to $c=0.9$ ($R^2 = 0.60$ for $X_{\text{pop}}$ and $R^2=0.19$ for $Y$), the PPI effective sample size only outperforms the GT-only method by $1.9\times$ for elevation and $1.3\times$ for population.    

As the bias level increases, the PPI intervals grow wider but remain significantly narrower than the GT-only interval, because the biased map (which uses a square root transformation of the true variable values) still contains significant information about the true population and forest cover values. For example, when the bias level is $c=0.2$ in both $X_{\text{pop}}$ and $Y$ ($R^2 = 0.997$ for both), the PPI effective sample size outperforms the GT-only method by a factor of $115\times$ for elevation and $85\times$ for population. When the bias level increases to $c=0.9$ ($R^2 = 0.94$ for both), the PPI effective sample size only outperforms the GT-only method by $20\times$ for elevation and $7\times$ for population.

\section{Discussion}
Using remote sensing-based maps to draw scientific inferences requires uncertainty quantification and propagation of uncertainty to downstream analyses --- otherwise, we risk making, in the words of \citet{mcroberts2011satellite}, just pretty pictures. In this work, we applied prediction-powered inference, using remote sensing map products along with a small amount of ground truth data, to estimate regression coefficients, while correctly taking into account the errors in the map products. We find that PPI estimates have narrower confidence intervals than the GT-only estimates. Unlike the map-only estimator using the map product alone, the PPI estimator is guaranteed to be unbiased.

\paragraph{Bias is the main issue for the map-only estimator.} Our experiments illustrate that a map-only regression coefficient estimate without bias correction can yield incorrect inferences. We see this clearly in our examples with MOSAIKS data in which we know the true coefficients. In Example 3 (housing price linear regression), the map-only estimator underestimates the intercept and overestimates the magnitude of the nightlight and road length coefficients. In Example 4 (forest cover linear regression), the map-only estimator underestimates the intercept, underestimates the magnitude of the elevation coefficient, and overestimates the magnitude of the population coefficient. In our map error simulations with forest cover linear regression (Section \ref{sec:simulations}), the map-only estimates of elevation and population coefficients become increasingly biased as we add noise in $X$ or add bias in $X$ or $Y$ to the simulated maps. In all cases, the map-only estimator results in very narrow confidence intervals that give a false sense of certainty. To address the bias of map-only estimators and the under-coverage of their confidence intervals, we use PPI to incorporate ground truth data into our estimates.

\paragraph{Performance of PPI.} PPI uses the ground truth calibration data to correct the map product, resulting in an unbiased regression coefficient estimator. Unlike the map-only estimator, the PPI confidence intervals are guaranteed to have the appropriate coverage probability (e.g. 95\%) for the true coefficient of interest. This is seen in Examples 3 and 4 (housing price and forest cover linear regressions), where the PPI confidence intervals contain the true coefficients for all covariates. In our map error simulations, all but one of the PPI confidence intervals contain the true coefficient, even when there is a high level of noise or bias in the map. 

The width of PPI confidence intervals compared to the GT-only method depends on the quality of the map product. If the map product is fairly accurate (i.e., the map product predictions $(\hat{X}, \hat{Y})$ are close to the ground truth values $(X, Y)$), the PPI regression coefficient confidence intervals are generally substantially narrower than the GT-only intervals. However, if the map product is low quality, the PPI estimator does not offer a significant advantage over the GT-only estimator. For instance, in our map noise simulations, the PPI confidence intervals become wider and approach the GT-only confidence intervals as noise increases and the quality of the map deteriorates. We also see in our experiments that the performance of PPI varies for different covariates. For instance, in Example 3, the PPI effective sample size outperforms the GT-only estimator by a factor of $16.7\times$ for the income covariate, while the road length covariate has a more modest $1.6\times$ improvement. 

\paragraph{Recommendations for map producers and users.} Ultimately, the best methods combine map products with ground truth to yield the narrowest intervals (highest certainty).
We recommend that producers of machine learning-generated remote sensing maps use a randomly sampled (i.e., via a probability sampling design) holdout ground truth dataset if they wish to estimate and construct confidence intervals for regression coefficients. The criterion for this holdout set is that the model, including during hyperparameter tuning, was not trained on it; otherwise, our confidence intervals may be too narrow (the prediction errors on the holdout set could be artificially small if the model is trained on these data points). A standard test set would be adequate, provided that it is a probability sample from the inference population. The holdout set can be obtained by simple random sampling or other probability sampling designs
such as unequal probability sampling, stratified random sampling, or cluster random sampling. 

Map producers should release this ground truth data and details of its sampling scheme alongside their map product to make their map suitable for downstream scientific inferences. When there are privacy concerns for releasing ground truth data, map producers can alternatively collaborate with map users on federated approaches for PPI, such as those described in \citet{MiaoLuNeurIPS}. In the absence of this, we recommend that map users generate their own ground truth data for the variable of interest if possible (e.g., through manual inspection of remote sensing imagery). 

\paragraph{Limitations.} PPI is limited to areas in which ground truth data are available. However, this would be necessary for any method that does not make further assumptions about the errors of the machine-learning model. Furthermore, the calibration dataset must be a probability sample from the ground-truth distribution of interest (this ensures that the PPI and GT-only estimators will both be unbiased). If ground truth data are obtained using non-probability sampling, the regression coefficient estimates using these ground truth data points may not reflect the actual relationships between variables in the population of interest.

Together, these attributes mean that such methods may be especially straightforward to use when ground truth data can be generated \emph{de novo}, such as from human inspection of available satellite imagery, and more challenging to use when ground truth data requires field data collection or survey data collection. 

The current theory for PPI focuses on the i.i.d. case, whereas in remote sensing we would usually have a fixed area of interest and then randomly sample a subset of the data points in the area as the calibration set (\emph{design-based inference}). In survey sampling it is known that the former often approximates the latter well when the sample is small relative to the total population, but admittedly this connection has not yet been made explicit for PPI. In addition, the current PPI approach for constructing confidence intervals does not account for possible spatial autocorrelations in the data, although future work can develop adjustments for spatial autocorrelations. Moreover, we suspect that in many applications accounting for spatial autocorrelations would only marginally increase the widths of the PPI confidence intervals. This is because the PPI estimator can be decomposed into the sum of two terms, both of which have variances that are unlikely to change much when accounting for short-range spatial autocorrelations (one term is based on a very large sample so it should have small variance regardless; the other term is based on a small calibration sample whose data points are likely to be far apart from each other).


While PPI applies to settings where the ground truth sample is collected using unequal probability sampling, stratified random sampling, or cluster random sampling~\cite{Kluger25GeneralizingPPI}, we note that this work does not explore these common settings. In addition, PPI can be used for a variety of regression models beyond generalized linear models such as quantile regression (e.g., \citet{Kluger25GeneralizingPPI,MiaoLuNeurIPS}), although due to space constraints we only consider logistic and linear regression models in this paper. 



\section*{Acknowledgments}
This material is based upon work supported by the U.S. Department of
Energy, Office of Science, Office of Advanced Scientific Computing Research, Department of
Energy Computational Science Graduate Fellowship under Award Number(s) DE-SC0023112. This report was prepared as an account of work sponsored by an agency of the
United States Government. Neither the United States Government nor any agency thereof, nor
any of their employees, makes any warranty, express or implied, or assumes any legal liability
or responsibility for the accuracy, completeness, or usefulness of any information, apparatus,
product, or process disclosed, or represents that its use would not infringe privately owned
rights. Reference herein to any specific commercial product, process, or service by trade name,
trademark, manufacturer, or otherwise does not necessarily constitute or imply its
endorsement, recommendation, or favoring by the United States Government or any agency
thereof. The views and opinions of authors expressed herein do not necessarily state or reflect
those of the United States Government or any agency thereof.

\printbibliography

\newpage
\listoffigures

\newpage
\begin{appendices}
\section{Superpopulation-based versus design-based inference}\label{appendix:SuperpopulationInferenceJustification}

This paper focuses on a superpopulation-based inference framework where the goal is to estimate a quantity that describes the broader population from which the sample was drawn. This inference framework is in contrast to the design-based inference framework \cite{sarndal1992model} that is commonly used in the remote sensing literature for area estimation tasks \cite{Stehman2000,olofsson2014good}. In the design-based inference framework the observed samples from the map data are presumed to constitute the entire population of interest. In this appendix, we clarify the differences between design-based inference and superpopulation-based inference and explain why we use the latter framework.

\subsection{Design-based inference framework}
In the \textbf{design-based inference} framework, the investigator's goal is to estimate a parameter that describes a finite, but large population. In particular, suppose there is a population of $N$ pixels with ground truth values of the response variables and covariates denoted by $(y_i,x_i)_{i=1}^N$ and map-based predictions of these variables denoted by $(\hat{y}_i,\hat{x}_i)_{i=1}^N$, all of which are observed. (Here, we use lowercase letters for $y_i$, $x_i$, $\hat{y}_i$, and $\hat{x}_i$ to emphasize that in design-based inference these quantities are viewed as fixed values rather than random variables, which are typically denoted with uppercase letters.) While the ground truth values $y_i$ and $x_i$ exist for every $i \in \{1,\dots,N\}$, they are only observed for a small calibration subsample of size $n$ out of the total population of size $N$. In design-based inference, the goal is to estimate a parameter that is an exact function of the data from the entire population $(y_i,x_i)_{i=1}^N$. For example, when using a design-based inference framework for mean estimation tasks, the goal is to estimate $$\theta_*^{\text{Des.}}=\frac{1}{N}\sum_{i=1}^N y_i,$$ or the average of $y$ across the population of interest. Similarly, in linear regression coefficient estimation tasks in a design-based inference framework, the goal would be to estimate $$\beta_*^{\text{Des.}}=\argmin\limits_{\beta \in \mathbb{R}^p} \Bigl\{ \frac{1}{N}\sum_{i=1}^N (y_i - \beta^\tran x_i)^2 \Bigr\},$$ which gives the least squares regression coefficient across the finite population of interest. Notably, $\theta_*^{\text{Des.}}$ and $\beta_*^{\text{Des.}}$ could be precisely calculated if all $y_i$ and $x_i$ were observed for all units in the population. However, the estimands $\theta_*^{\text{Des.}}$ and $\beta_*^{\text{Des.}}$ are unknown because the $y_i$ and $x_i$ values are unobserved for most $i \in \{1,\dots,N\}$. Instead, $\theta_*^{\text{Des.}}$ or $\beta_*^{\text{Des.}}$ must be estimated from the available data. Let $\hat{\theta}_*^{\text{Des.}}$ and $\hat{\beta}_*^{\text{Des.}}$ denote estimators of $\theta_*^{\text{Des.}}$ or $\beta_*^{\text{Des.}}$ from the observed data, respectively. In design-based inference, when calculating the standard errors of $\hat{\theta}_*^{\text{Des.}}$ and $\hat{\beta}_*^{\text{Des.}}$, the data is assumed to be nonrandom and fixed. Instead all of the randomness in $\hat{\theta}_*^{\text{Des.}}$ and $\hat{\beta}_*^{\text{Des.}}$ is driven by the random selection of a subsample of size $n$ to be labelled (i.e, to have $y_i$ and $x_i$ observed) out of the population of size $N$. 

\subsection{Superpopulation-based inference}

In the \textbf{superpopulation-based inference} framework, the investigator's goal is to estimate a parameter that describes a very large (or infinite population) from which the data was drawn. In particular, let $Y$ and $X$ be random variables denoting the response variable and covariates of interest and let $\hat{Y}$ and $\hat{X}$ denote remotely sensed predictions of these variables. Let $\mathbb{P}$ denote the joint probability distribution of the random vector $(Y,X,\hat{Y},\hat{X})$. 
In superpopulation-based inference, the assumption is that data consists of a sample of size $N$, denoted by $(Y_i,X_i,\hat{Y}_i,\hat{X}_i)_{i=1}^N$, where for each $i$, $(Y_i,X_i,\hat{Y}_i,\hat{X}_i)$ is a random draw from a probability distribution $\mathbb{P}$. (As before, $Y_i$ and $X_i$ are only observed on a random subsample of size $n$ rather than the larger sample of size $N$, but in principle $Y_i$ and $X_i$ still exist for each $i \in \{1,\dots,N\}$).

In superpopulation-based inference the goal is to estimate some parameter describing the distribution $\mathbb{P}$ from which each data point $(Y_i,X_i,\hat{Y}_i,\hat{X}_i)$ is assumed to be drawn. For example, when using a superpopulation-based inference framework for mean estimation tasks, the goal is to estimate $$\theta_*^{\text{Sup.}}=\e[Y]$$ or the expected value of $Y$ with respect to the distribution $\mathbb{P}$. Similarly, in regression coefficient estimation tasks in a design-based inference framework, the goal would be to estimate $$\beta_*^{\text{Sup.}}=\argmin\limits_{\beta \in \mathbb{R}^p} \Bigl\{ \e \big[ (Y - \beta^\tran X)^2 \big] \Bigr\},$$ which gives the coefficient vector minimizing the expected squared error. Notably, $\theta_*^{\text{Sup.}}$ and $\beta_*^{\text{Sup.}}$ cannot be calculated exactly even if all $Y_i$ and $X_i$ are observed on all $N$ sampled data points drawn from $\mathbb{P}$. 

We remark that the distribution $\mathbb{P}$ from which each data point $(Y_i,X_i,\hat{Y}_i,\hat{X}_i)$  is assumed to be drawn need not be known to the investigator and could be completely arbitrary. $\mathbb{P}$ is not restricted to a parametric class of distributions and no specific modelling assumptions about the distribution $\mathbb{P}$ are needed. We therefore use the terminology superpopulation-based inference rather than ``model-based inference" (which is also often used), because the latter terminology would misleadingly suggest that some type of parametric modelling assumptions are needed. 

\subsection{Why we use superpopulation-based inference}

 

 We focus on superpopulation-based inference approaches as opposed to design-based inference approaches for the following three reasons: 

 \paragraph{1) Superpopulation-based inference is a suitable framework when the goal is estimating regression coefficients to study causal or associative relationships between variables.}

 When the goal is estimating the area of a certain land class over a fixed geographical region, using a design-based inference approach is considered a ``good practice" \cite{olofsson2014good}. In contrast, when the investigator wishes to estimate a regression coefficient rather than an area, superpopulation-based inference could be more appropriate. As noted in \cite{Stehman2000}, a useful guideline for determining whether to use a design-based or superpopulation-based framework would be to ask the question of whether one would be able to answer the research question of interest with certainty if ground truth data were available for all $N$ sampled units. If not, they state ``the inferential objective is likely a process or more expansive superpopulation generalizing beyond the $N$ pixels of the observable population, and [superpopulation-based] inference applies." We focus on settings where the goal is to estimate a regression coefficient in order to study a causal or associative relationship between variables. In these settings of interest, superpopulation-based inference is a suitable framework because the goal is to quantify a parameter describing a phenomenon that generalizes beyond the available observations from the map product.

 \paragraph{2) Existing methods for our use cases of interest use a superpopulation-based inference framework.}
The Prediction-Powered Inference (PPI) literature has thus far only developed methods that are theoretically justified in superpopulation-based inference settings \cite{angelopoulos2023prediction,angelopoulos2023ppi++,Kluger25GeneralizingPPI}. This is not a fundamental limitation of PPI methods, and future work can extend such methods to design-based inference settings. 
To our knowledge, in settings where the goal is to estimate a regression coefficient vector, there are no existing methods that use a design-based inference framework and also leverage both remotely sensed maps and a calibration sample with ground truth measurements. (For area estimation tasks, such approaches exist in a design-based inference framework \cite{olofsson2014good}).

\paragraph{3) In many motivating applications a design-based inference framework would yield similar estimates and confidence intervals.}

It is common in remote sensing studies to downsample the data from a remote sensing map to save computational resources, especially when the goal is to fit a complex, computationally expensive regression model. For example, \cite{ZhouEtAlCropRotationDownsamples} downsamples a crop type and crop yield map to fit a model relating crop rotation to crop yield for computational reasons (they find little change in their results across 20 different random downsamplings of the map products). Similarly, in the examples we study in this manuscript (Section \ref{sec:examples}), we use a downsampling of the original map product rather than the entire map product. In these settings where the remote sensing map is downsampled, design-based inference approaches and superpopulation-based inference approaches would give nearly identical estimates and confidence intervals. In particular, if a map has $M$ pixels and the user downsamples the map and only considers $N \ll M$ pixels, applying a design-based inference approach to the finite population of $M$ pixels would yield nearly identical confidence intervals to applying a superpopulation-based inference approach that assumes the data from the $N$ pixels are each draws from an unknown probability distribution $\mathbb{P}$ (or superpopulation). Therefore, in many applications of interest design-based inference and superpopulation-based inference approaches would yield similar estimates and confidence intervals.
     
Even if no downsampling occurs, the results from design-based inference and superpopulation-based inference would be similar provided that $N$, the number of map-based observations, is large and much larger than $n$, the number of calibration samples. In particular, recall that the PPI estimator has the form $$\hat{\beta}_{\text{PPI}} = \hat{\Omega} \hat{\gamma}_{\text{map}} + (\hat{\beta}_{\text{calib}}-\hat{\Omega} \hat{\gamma}_{\text{calib}}), $$ for some well chosen $\hat{\Omega} \in \mathbb{R}^{d \times d}$. If $N$ is large and no downsampling occurs, $\hat{\gamma}_{\text{map}}$ will have zero variance in a design-based inference framework and will have very small variance in a superpopulation-based inference framework. Meanwhile, if $N$ is large and $n \ll N$, the variances of the $\hat{\beta}_{\text{calib}}$ and $\hat{\gamma}_{\text{calib}}$ terms will be similar whether one uses a design-based inference framework or a superpopulation-based inference framework. As a result, we expect that both frameworks will yield confidence intervals of similar widths for the above estimator $\hat{\beta}_{\text{PPI}}$, with the design-based inference confidence intervals being slightly narrower and the differences diminishing in the limit as $N \to \infty$ and $n/N \to 0$.
    

\newpage
\section{Formulas for variance and tuning matrix of $\hat{\beta}_{\text{PPI}}$}\label{appendix:FormulaForPPIoptTuningAndCov}

In this appendix, we present formulas for the covariance matrix of the proposed estimator $\hat{\beta}_{\text{PPI}}$ and its optimal tuning matrix. We focus on providing explicit formulas in the case of linear regression and logistic regression. The resulting formulas can be found in \cite{chen2000unified}, but for ease of exposition, we provide a slightly different derivation and more explicit expressions in our cases of interest. More formal derivations in a more general setting can be found in the appendix of \cite{Kluger25GeneralizingPPI}.

Recall that $\hat{\gamma}_{\text{map}}$, $\hat{\beta}_{\text{calib}}$, and $\hat{\gamma}_{\text{calib}}$ are regression coefficient estimators from a generalized linear model (GLM), which could be a linear model or logistic regression model. Moreover, $\hat{\gamma}_{\text{map}}$ is estimated using a sample $(\hat{X}_i',\hat{Y}_i')_{i=1}^N$ of $N$ map-based predictions. Meanwhile, $\hat{\beta}_{\text{calib}}$ and $\hat{\gamma}_{\text{calib}}$ are estimated using a simple random subsample of size $n$, which is a fraction $\rho=n/N$ of the size of the sample of map-based predictions. $\hat{\beta}_{\text{calib}}$ is estimated using ground truth data points $(X_i,Y_i)_{i=1}^n$ and  $\hat{\gamma}_{\text{calib}}$ is estimated using the corresponding proxies $(\hat{X}_i,\hat{Y}_i)_{i=1}^n$. For convenience we let $(X,Y)$ denote a random vector with the same distribution as $(X_i,Y_i)$ and $(\hat{X},\hat{Y})$ denote a random vector with the same distribution as the map-based predictions $(\hat{X}_i,\hat{Y}_i)$ or $(\hat{X}_i',\hat{Y}_i')$. 

Now let $\psi : \mathbb{R} \to \mathbb{R}$ be a function that specifies the type of GLM being run: $\psi(s)=\frac{1}{2}s^2$ for linear regression and $\psi(s)=\log(1+e^s)$ for logistic regression with a canonical link function. Note since $\hat{\beta}_{\text{calib}}$ is a regression coefficient from a GLM, it estimates the well-defined quantity $$\beta_*= \argmin_{\beta \in \mathbb{R}^p} \e[-Y \beta^\tran X + \psi (\beta^\tran X)].$$ Similarly, $\hat{\gamma}_{\text{calib}}$ and $\hat{\gamma}_{\text{map}}$ estimate the well defined quantity $$\gamma_*= \argmin_{\beta \in \mathbb{R}^p} \e[-\hat{Y} \beta^\tran \hat{X} + \psi (\beta^\tran \hat{X})].$$

 Based on asymptotic theory for M-estimators (e.g., Chapter 5 of \cite{VanderVaartTextbook}) and since GLM estimators are M-estimators, under standard regularity conditions the 3 component estimators of $\hat{\beta}_{\text{PPI}}$ are asymptotically multivariate normal with $$\begin{bmatrix} \hat{\beta}_{\text{calib}} \\ \hat{\gamma}_{\text{calib}} \\ \hat{\gamma}_{\text{map}} \end{bmatrix} \stackrel{\cdot}{\sim} \mathcal{N} \Bigg( \begin{bmatrix} \beta_* \\ \gamma_* \\ \gamma_* \end{bmatrix}, \frac{1}{n} \begin{bmatrix}
     D_{\beta}^{-1} C_{11} D_{\beta}^{-1} & D_{\beta}^{-1} C_{12} D_{\gamma}^{-1} & \rho D_{\beta}^{-1} C_{12} D_{\gamma}^{-1} \\ D_{\gamma}^{-1} C_{12}^\tran D_{\beta}^{-1} & D_{\gamma}^{-1} C_{22} D_{\gamma}^{-1} & \rho D_{\gamma}^{-1} C_{22} D_{\gamma}^{-1}
    \\ \rho D_{\gamma}^{-1} C_{12}^\tran D_{\beta}^{-1} & \rho D_{\gamma}^{-1} C_{22} D_{\gamma}^{-1} & \rho D_{\gamma}^{-1} C_{22} D_{\gamma}^{-1}
\end{bmatrix} \Bigg).$$ Above, $\stackrel{\cdot}{\sim}$ means approximately distributed as, $\mathcal{N}(\mu,\Sigma)$ denotes a multivariate normal distribution with mean vector $\mu$ and covariance matrix $\Sigma$, and the component matrices are given by $$D_{\beta} = \mathbb{E}[\ddot{\psi}(\beta_*^\tran X) X X^\tran ], \quad D_{\gamma} = \mathbb{E}[\ddot{\psi}(\gamma_*^\tran \hat{X}) \hat{X} \hat{X}^\tran ], \quad C_{11}=\mathbb{E} \big[ \big(Y-\dot{\psi}(\beta_*^\tran X) \big)^2 X X^\tran \big]$$ $$C_{12}=\mathbb{E} \big[ \big(Y-\dot{\psi}(\beta_*^\tran X) \big) \big(\hat{Y}-\dot{\psi}(\gamma_*^\tran \hat{X}) \big) X \hat{X}^\tran \big] \quad \text{and} \quad C_{22}=\mathbb{E} \big[ \big(\hat{Y}-\dot{\psi}(\gamma_*^\tran \hat{X}) \big)^2 \hat{X} \hat{X}^\tran \big],$$ where $\dot{\psi}(\cdot)$ and $\ddot{\psi}(\cdot)$ give the first and second derivatives of the GLM class-specific function $\psi$. Now recall $\hat{\beta}_{\text{PPI}}$ takes the form

$$\hat{\beta}_{\text{PPI},\Omega} = \Omega \hat{\gamma}_{\text{map}} +(\hat{\beta}_{\text{calib}} - \Omega \hat{\gamma}_{\text{calib}}),$$ where $\Omega \in \mathbb{R}^{p \times p}$ is some tuning matrix. Note that by the previous asymptotic normality approximation and the formula for the distribution of a linear transformation of a multivariate Gaussian (i.e., using the fact that $A Z \sim \mathcal{N}( A\mu,A\Sigma A^\tran)$ when $A$ is a fixed matrix and $Z \sim \mathcal{N}(\mu,\Sigma)$),

$$\hat{\beta}_{\text{PPI},\Omega}=\begin{bmatrix}
    I_{p \times p} & -\Omega & \Omega
\end{bmatrix} \begin{bmatrix} \hat{\beta}_{\text{calib}} \\ \hat{\gamma}_{\text{calib}} \\ \hat{\gamma}_{\text{map}} \end{bmatrix} \stackrel{\cdot}{\sim} \mathcal{N}\Big( \beta_*, \frac{1}{n} V(\Omega) \Big),$$ where $$V(\Omega)= D_{\beta}^{-1} C_{11} D_{\beta}^{-1} -(1-\rho) \Omega [D_{\beta}^{-1} C_{12} D_{\gamma}^{-1} ]^\tran-(1-\rho)  D_{\beta}^{-1} C_{12} D_{\gamma}^{-1} \Omega^\tran + (1-\rho) \Omega D_{\gamma}^{-1} C_{22} D_{\gamma}^{-1} \Omega^\tran.$$ Note that the asymptotic variance of each component of $\hat{\beta}_{\text{PPI}}$ is minimized when the diagonal entries of $V(\Omega)$ are minimized. Since $V(\Omega)$ is quadratic in $\Omega$ with the $j$th diagonal entry only depending on the $j$th row of $\Omega$, an exercise in linear algebra and multivariate calculus for quadratic functions shows that choosing $$\Omega_*= D_{\beta}^{-1} C_{12} C_{22}^{-1} D_{\gamma},$$ simultaneously minimizes each diagonal component of $V(\Omega)$. Thus $\Omega_*$ is the optimal tuning matrix, and note that it is identical to the tuning matrix used in Equation (4) of \cite{chen2000unified}. Further note that $$V(\Omega_*)=D_{\beta}^{-1} C_{11} D_{\beta}^{-1} -(1-\rho)D_{\beta}^{-1} C_{12} C_{22}^{-1} C_{12}^\tran D_{\beta}^{-1},$$ which matches the asymptotic variance found in Equation (5) of \cite{chen2000unified}.

If $\hat{D}_{\beta}$, $\hat{D}_{\gamma}$, $\hat{C}_{11}$, $\hat{C}_{12}$ and $\hat{C}_{22}$ are data-based estimators of previously defined matrices that converge in probability to $D_{\beta}$, $D_{\gamma}$, $C_{11}$, $C_{12}$, and $C_{22}$, respectively, then choosing the tuning matrix \begin{equation}\label{eq:OptOmega}
 \hat{\Omega}=\hat{D}_{\beta}^{-1} \hat{C}_{12} \hat{C}_{22}^{-1} \hat{D}_{\gamma}
\end{equation} will be asymptotically optimal. In such settings $\hat{\Omega}$ will converge in probability to $\Omega_*$ and asymptotically, $\hat{\beta}_{\text{PPI}}$ will approximately follow a $\mathcal{N}\big(\beta_*,  \frac{1}{n} V(\Omega_*) \big)$ distribution, in which case one can use an approximated covariance matrix for $\hat{\beta}_{\text{PPI}}$ given by \begin{equation}\label{eq:EstimatedCovMat}
    \hat{V}_{\text{PPI}}=\frac{1}{n} \Big( \hat{D}_{\beta}^{-1} \hat{C}_{11} \hat{D}_{\beta}^{-1} -
\big(1-\frac{n}{N} \big)\hat{D}_{\beta}^{-1} \hat{C}_{12} \hat{C}_{22}^{-1} \hat{C}_{12}^\tran \hat{D}_{\beta}^{-1} \Big)
\end{equation} and the normal approximation to calculate standard errors and confidence intervals for $\hat{\beta}_{\text{PPI}}$. Note that in this paper we use a bootstrap approach to constructing confidence intervals because it can generalize to other settings without having to find an explicit formula for $\hat{V}_{\text{PPI}}$.

\subsection{Explicit formulas in linear regression settings}

In linear regression settings $\psi(s)=\frac{1}{2} s^2$, so $\dot{\psi}(s)=s$ and $\ddot{\psi}(s)=1.$ Hence $D_{\beta}=\e[XX^\tran]$, $D_{\gamma}=\e[\hat{X}\hat{X}^\tran]$, $$C_{11}=\mathbb{E} \big[ (Y-\beta_*^\tran X)^2 X X^\tran \big], C_{12}=\mathbb{E} \big[ (Y-\beta_*^\tran X) (\hat{Y}-\gamma_*^\tran \hat{X})  X \hat{X}^\tran \big], \quad \text{and} \quad C_{22}=\mathbb{E} \big[ (\hat{Y}-\gamma_*^\tran \hat{X} )^2 \hat{X} \hat{X}^\tran \big].$$ Thus one can use the plug in estimators $$\hat{D}_{\beta}= \frac{1}{n} \sum_{i=1}^n X_i X_i^\tran, \quad \hat{D}_{\gamma}= \frac{1}{n} \sum_{i=1}^n \hat{X}_i \hat{X}_i^\tran, \quad \hat{C}_{11}= \frac{1}{n} \sum_{i=1}^n (Y_i-\hat{\beta}_{\text{calib}}^\tran X_i)^2 X_i X_i^\tran,$$ $$\hat{C}_{12}=\frac{1}{n} \sum_{i=1}^n (Y_i-\hat{\beta}_{\text{calib}}^\tran X_i) (\hat{Y}_i-\hat{\gamma}_{\text{calib}}^\tran \hat{X}_i)  X_i \hat{X}_i^\tran, \quad \text{and } \hat{C}_{22}=\frac{1}{n} \sum_{i=1}^n (\hat{Y}_i-\hat{\gamma}_{\text{calib}}^\tran \hat{X}_i )^2 \hat{X}_i \hat{X}_i^\tran.$$ One can then plug the above formulas into Equation \eqref{eq:OptOmega} to obtain an asymptotically optimal tuning matrix choice. They can further plug these formulas into Equation \eqref{eq:EstimatedCovMat} to estimate the variance of $\hat{\beta}_{\text{PPI}}$ with this tuning matrix choice (the estimated variance matrix of $\hat{\beta}_{\text{PPI}}$ can then be used to calculate standard errors and confidence intervals).

\subsection{Explicit formulas in logistic regression settings}

In logistic regression settings $\psi(s)=\log(1+e^s)$, so $\dot{\psi}(s)=1/(1+e^{-s})$ and $\ddot{\psi}(s)=e^s/(1+e^{s})^2$. We can then similarly use the plug in estimators for $D_{\beta}$, $D_{\gamma}$, $C_{11}$, $C_{22}$, and $C_{12}$ given by $$\hat{D}_{\beta}= \frac{1}{n} \sum_{i=1}^n \frac{\exp(\hat{\beta}_{\text{calib}}^\tran X_i)}{\big(1+\exp(\hat{\beta}_{\text{calib}}^\tran X_i) \big)^2} X_i X_i^\tran, \quad \hat{D}_{\gamma}= \frac{1}{n} \sum_{i=1}^n \frac{\exp(\hat{\gamma}_{\text{calib}}^\tran \hat{X}_i)}{\big(1+\exp(\hat{\gamma}_{\text{calib}}^\tran \hat{X}_i) \big)^2}  \hat{X}_i \hat{X}_i^\tran,$$  $$ \hat{C}_{11}= \frac{1}{n} \sum_{i=1}^n \Big(Y_i-\frac{1}{1+\exp(-\hat{\beta}_{\text{calib}}^\tran X_i)} \Big)^2 X_i X_i^\tran, \quad \hat{C}_{22}= \frac{1}{n} \sum_{i=1}^n \Big(\hat{Y}_i-\frac{1}{1+\exp(-\hat{\gamma}_{\text{calib}}^\tran \hat{X}_i)} \Big)^2 \hat{X}_i \hat{X}_i^\tran$$ $$\text{and} \quad \hat{C}_{12}=\frac{1}{n} \sum_{i=1}^n \Big(Y_i-\frac{1}{1+\exp(-\hat{\beta}_{\text{calib}}^\tran X_i)} \Big) \Big(\hat{Y}_i-\frac{1}{1+\exp(-\hat{\gamma}_{\text{calib}}^\tran \hat{X}_i)} \Big) X_i \hat{X}_i^\tran.$$ One can then plug the above formulas into Equation \eqref{eq:OptOmega} to obtain an asymptotically optimal tuning matrix choice. They can further plug these formulas into Equation \eqref{eq:EstimatedCovMat} to estimate the variance of $\hat{\beta}_{\text{PPI}}$ with this tuning matrix choice (the estimated variance matrix of $\hat{\beta}_{\text{PPI}}$ can then be used to calculate standard errors and confidence intervals).

\newpage
\section{Datasets}\label{appendix:datasets}
We describe the datasets used in the four examples in the paper. 

\subsection{Example 1: Deforestation logistic regression}\label{appendix:datasets-deforestation}
Each data point is a pixel with 30 m resolution. For the deforestation response variable, we use $N=963548$ map product data points selected via simple random sampling from the NASA Global Forest Cover Change (GFCC) map \cite{townshend2016global} in the study area. The GFCC product is derived from Landsat 5 and Landsat 7 surface reflectance imagery, and includes maps for tree cover percentage in each pixel in five-year increments from 2000 to 2015. 

For ground truth, we use a Amazon deforestation and forest disturbance dataset created by \cite{bullock2020satellite} using time-series analysis of Landsat imagery and high-resolution imagery from Google Earth between 1995 and 2017. The dataset records the years corresponding to deforestation, degradation, and natural disturbance events at each ground truth data point. We use this to construct a binary ground truth dataset restricted to Brazil, in which we consider a data point deforested if it experienced a deforestation event between 2000 and 2015. The original dataset is obtained by stratified random sampling, using a stratification map (also created by \cite{bullock2020satellite}) that contains 7 strata related to deforestation status. For our experiments, we construct a ``simple random sample" ground truth dataset of size $n=1386$ by downsampling the data points in each of the strata to match its area proportion. Using the ground truth dataset, we find that for the ``deforested" class, the remotely sensed map has 69\% producer's accuracy and 71\% user's accuracy. 

\begin{figure}[h]
\centering
\includegraphics[width=0.4\textwidth]{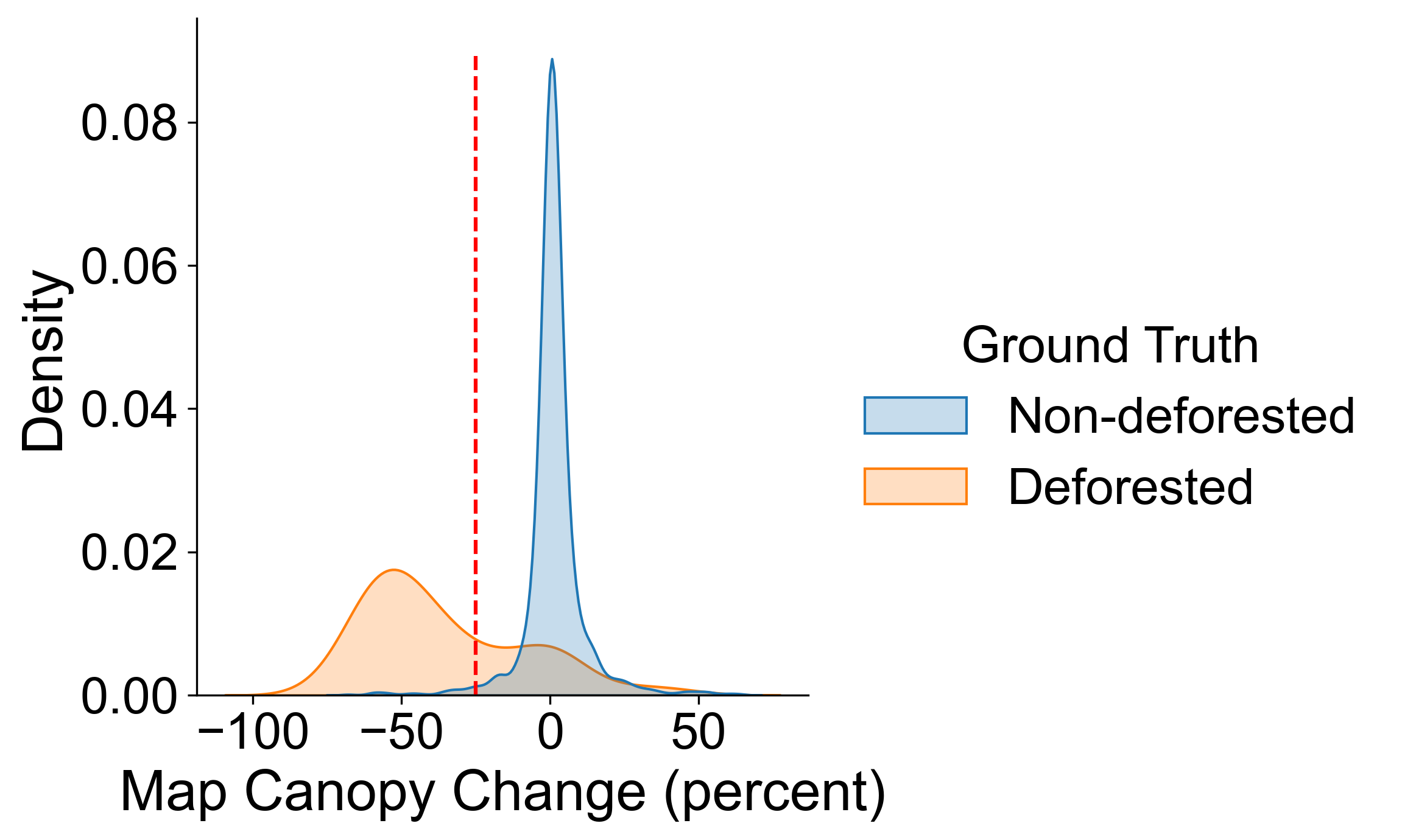}

\caption{\textbf{NASA GFCC map canopy cover change (2000-2015) distributions for non-deforested and deforested ground truth} over a simple random sample of $n=1386$ data points. The distributions are visualized using kernel density estimation. We choose a -25\% deforestation threshold (shown in red) to binarize the map product.}\label{deforestation-binarization}

\end{figure}

\paragraph{Binarizing the NASA GFCC map product for deforestation.} The ground truth data points from \cite{bullock2020satellite} have binary deforestation values ($Y=1$ if deforested and $Y=0$ if not deforested between 2000-2015), while the NASA GFCC map product gives canopy cover percentages over time at each pixel. In order to binarize the continuous map product data, we classify a map pixel as ``deforested" if its canopy cover percentage satisfies $\text{canopy}_{2015} - \text{canopy}_{2000} \leq -25\%$. We choose the $-25\%$ threshold by visually comparing the map-derived canopy cover change distributions for deforested and non-deforested ground truth data points (Figure \ref{deforestation-binarization}). Most of the deforested ground truth data points have a map canopy cover decrease of more than $25\%$, while most of the non-deforested ground truth data points have either a map canopy cover increase, no map canopy cover change, or a map canopy cover decrease of less than $25\%$.

For the covariates, we use a map of Brazilian federal roads in 2000 \cite{de2021transport} and the WWF HydroSHEDS Free Flowing Rivers Network map \cite{grill2019mapping} (we include all rivers of order at most 3) to compute the distance from each data point to the nearest major road and major river. We obtain elevation and slope from the NASA Digital Elevation Model \cite{jpl2020} at 30 m resolution. Prior work has shown that deforestation in the Brazilian Amazon is more likely to occur near major roads and navigable rivers due to human settlement and economic activities along these transportation networks \cite{laurance2002predictors, kirby2006future, rosa2013predictive}. Elevation and slope have been associated with forest cover change; for example, deforestation is generally less likely on sloped terrain \cite{sandel2013human}. Studies have also regressed remotely sensed deforestation against other covariates such as climate, soil fertility, population, and protected area status \cite{etter2006regional, laurance2002predictors}. Additionally, there exist unofficial, non government built roads in Brazil that are not identified in the federal roads map but may be associated with deforestation \cite{arima2016explaining, brandao2006mapping, perz2007unofficial, valle2024automated}. However, in this paper, we do not use a comprehensive list of covariates that could influence deforestation, because the primary purpose of this regression is to illustrate the methods rather than explain deforestation.

\subsection{Example 2: Tree cover linear regression}\label{appendix:datasets-tree-cover}
Each data point is a pixel with 30 m resolution. For the tree cover response variable, we use $N=983238$ map product data points selected via simple random sampling from the 2021 USFS Tree Canopy Cover (TCC) product \cite{forest2023usfs}. The TCC product is derived from Landsat and Sentinel-2 top-of-atmosphere reflectance data, USDA NASS CDL data, and terrain data. Its tree cover percentage predictions are obtained from random forest models trained on reference data from human-labeled high-resolution imagery. For ground truth, we manually label $n=983$ data points by downloading high-resolution satellite images from Google Maps for tree cover percentage and visually estimating the proportion of each 30 m by 30 m image that is covered by trees. The image locations were selected via simple random sampling over the study area. The $R^2$ coefficient between the remotely sensed and ground truth tree cover percentage values is 0.68.

For the covariates, we use a map of the average Global Aridity Index from 1970-2000 (at 30 arc-seconds resolution) \cite{trabucco2019global} and the NASA Digital Elevation Model \cite{jpl2020} (at 30 m resolution). The aridity index is defined as a ratio between precipitation and evapo-transpiration, and is \textit{lower} for more arid areas; its value ranges from 0.02 to 6.55 in our dataset. Increased aridity is associated with lower tree cover (and decreased vegetation in general) \cite{bastin2017extent, williams2010forest, berdugo2020global}. Elevation and slope have also been associated with tree cover globally; there is generally more tree cover at lower elevations (e.g., due to higher temperatures) \cite{mayor2017elevation} and on sloped terrain \cite{sandel2013human}. We do not include a comprehensive list of covariates that could influence tree cover, because the primary purpose of this regression is to illustrate the methods rather than explain tree cover.

\subsection{Example 3: Housing price linear regression}\label{appendix:datasets-housing-price}
For all variables, we use the MOSAIKS dataset \cite{rolf2021generalizable} which contains an unequal probability sample of data points in the United States, where each data point is a grid cell with roughly 1 km resolution. (Specifically, the data points are sampled using an unequal probability sampling design, such that data points at locations with higher population density are more likely to be selected.) The dataset contains ground truth values across seven variables, as well as corresponding remotely sensed predictions obtained from a machine learning model trained on a subset of the ground truth data points. The model runs ridge regression on featurized representations of Google Static Maps satellite imagery. The MOSAIKS predictions generally capture the variables well, with $R^2$ ranging from 0.45 to 0.91 to depending on the variable. We note that \cite{proctor2023parameter} also runs regressions on the MOSAIKS dataset using the ground truth and remotely sensed predictions, but unlike PPI, their coefficient estimation method relies on Bayesian modelling assumptions and does not provide frequentest statistical guarantees for e.g., 95\% coverage of the true coefficients. 

The ground truth datasets are described in detail in the Supplementary Materials of \cite{rolf2021generalizable}. Housing price per square foot for sales since 2010 are obtained using the Zillow Transaction and Assessment Dataset. Median household income at the census block group level is obtained from the 2015 American Community Survey. Nighttime light intensity is obtained from average radiance values from the 2015 Visible Infrared Imaging Radiometer Suite (VIIRS). (Note that the nightlights data are based on nighttime remote sensing data rather than collected on the ground, but they can be viewed as ground truth relative to the MOSAIKS nightlights predictions, which are based on daytime satellite imagery.) Total road length in each $\sim$1 km grid cell is computed from the United States Geological Survey National Transportation Dataset. 

After removing data points with missing variables, there are $N=46418$ data points that have both ground truth and remotely sensed values for housing price (log dollars per square foot), household income (dollars), nighttime light intensity ($\log(1 + \text{nW}/\text{cm}^2/\text{sr})$), and road length (km). The $R^2$ coefficient between MOSAIKS predictions and ground truth is 0.85 for nightlights and 0.52 for road length.

For housing price and income, we use the ground truth values for all $46418$ data points. For nightlights and road length, we take a simple random subsample of $n=500$ data points as the ground truth calibration set (and use remotely sensed values for the remaining data points). For the GT-only method, we use only the $n=500$ ground truth data points for all four variables. We also compare the regression coefficient point estimates from our methods against the ``true coefficients" computed by running linear regression on the full set of $46418$ ground truth data points for all four variables. 

(Note that in Example 1, we treat observations of ``distance from roads" as ground truth because we are considering the map of Brazilian federal roads to be fairly accurate. By contrast, the remotely sensed road length data in Example 3 are obtained by training a machine learning model to predict road length from satellite imagery, so we expect this data to be less reliable.)

\subsection{Example 4: Forest cover linear regression}\label{appendix:datasets-forest-cover}
Similar to the previous example, we use the MOSAIKS dataset \cite{rolf2021generalizable}. The dataset also provides a simple random sample of data points in the United States, where each data point is a grid cell with roughly 1 km resolution. After removing data points with missing variables, there are $N=67968$ data points that have both ground truth and remotely sensed values for forest cover percentage, elevation (km), and population ($\log(1+\text{people})/\text{km}^2$). 

The ground truth datasets are described in detail in the Supplementary Materials of \cite{rolf2021generalizable}. Forest cover percentage is obtained from the 2010 Hansen Global Forest Change maps \cite{hansen2013high}, which are derived from Landsat imagery. (For illustrative purposes, we treat the Hansen forest cover values as ground truth and the MOSAIKS predictions as remotely sensed proxies.) Elevation is obtained from Mapzen terrain tiles with data from NASA Jet Propulsion Laboratory's Shuttle Radar Topography Mission. Population density is obtained from the Gridded Population of the World dataset, which uses the 2010 US Population and Housing Census. 

As in the previous example, the $N=67968$ remotely sensed predictions were obtained by \cite{rolf2021generalizable} using the MOSAIKS prediction model, which runs ridge regression on featurized representations of Google Static Maps satellite imagery. The $R^2$ coefficient between the MOSAIKS predictions and ground truth is 0.91 for forest cover and 0.73 for population.

For elevation, we use the ground truth values for all $67968$ data points. For forest cover and population, we take a simple random subsample of $n=500$ data points as the ground truth calibration set (and use remotely sensed values for the remaining data points). For the GT-only method, we use only the $n=500$ ground truth data points for all three variables. We also compare the regression coefficient point estimates from our methods against the ``true coefficients" computed by running linear regression on the full set of $67968$ ground truth data points for all four variables. 

To generate the histogram in the right panel of Figure \ref{motivating-example} of the main text, we compute PPI and GT-only coefficient estimates using 100 different simple random subsamples of $n=500$ ground truth data points from the full set of $67968$ ground truth data points. We use the same set of $N=67968$ map product data points in all 100 trials. Note that the map-only estimate remains the same in all trials. 

\newpage
\section{Simulating mean reversion bias in maps}\label{appendix:mean-reversion-bias}

In Section \ref{sec:simulations}, we simulate adding noise and nonlinear bias to the map products from Example 4 (forest cover linear regression). In this appendix, we simulate another model of map bias: linear mean reversion. Under this type of bias, the map underestimates high ground truth values and overestimates low ground truth values.

Using the same MOSAIKS dataset as Section \ref{sec:simulations}, we simulate different levels of mean reversion bias in the population map (error-in-$X$), the forest cover map (error-in-$Y$), or both (error-in-both). The elevation dataset is not altered. We compare the resulting 95\% coefficient confidence intervals at different levels of simulated map bias using map-only, GT-only, and PPI methods. 

\paragraph{Bias} To simulate bias for the ground truth population covariate $X_{\text{pop}}$, we linearly interpolate between $X_{\text{pop}}$ and the mean $\mu_{X_{\text{pop}}}$ of all $X_{\text{pop}}$ values in the full ground truth dataset (67968 data points). The resulting simulated map product value at ``bias level $c$" is 
\begin{align*}
    X_{\text{pop}}^{\text{map}} = c \mu_{X_{\text{pop}}} + (1-c) X_{\text{pop}}.
\end{align*}
Similarly, for the forest cover response variable $Y$ with mean $\mu_Y$, the simulated map product value at each data point is
\begin{align*}
    Y^{\text{map}} = c \mu_Y + (1-c) Y.
\end{align*}
We experiment with bias levels $c$ from 0 to 0.8 at increments of 0.1. When $c=0$, the simulated map is the same as the full ground truth map. 

\paragraph{Results}

\begin{figure}[t]
\centering
\includegraphics[width=\textwidth]{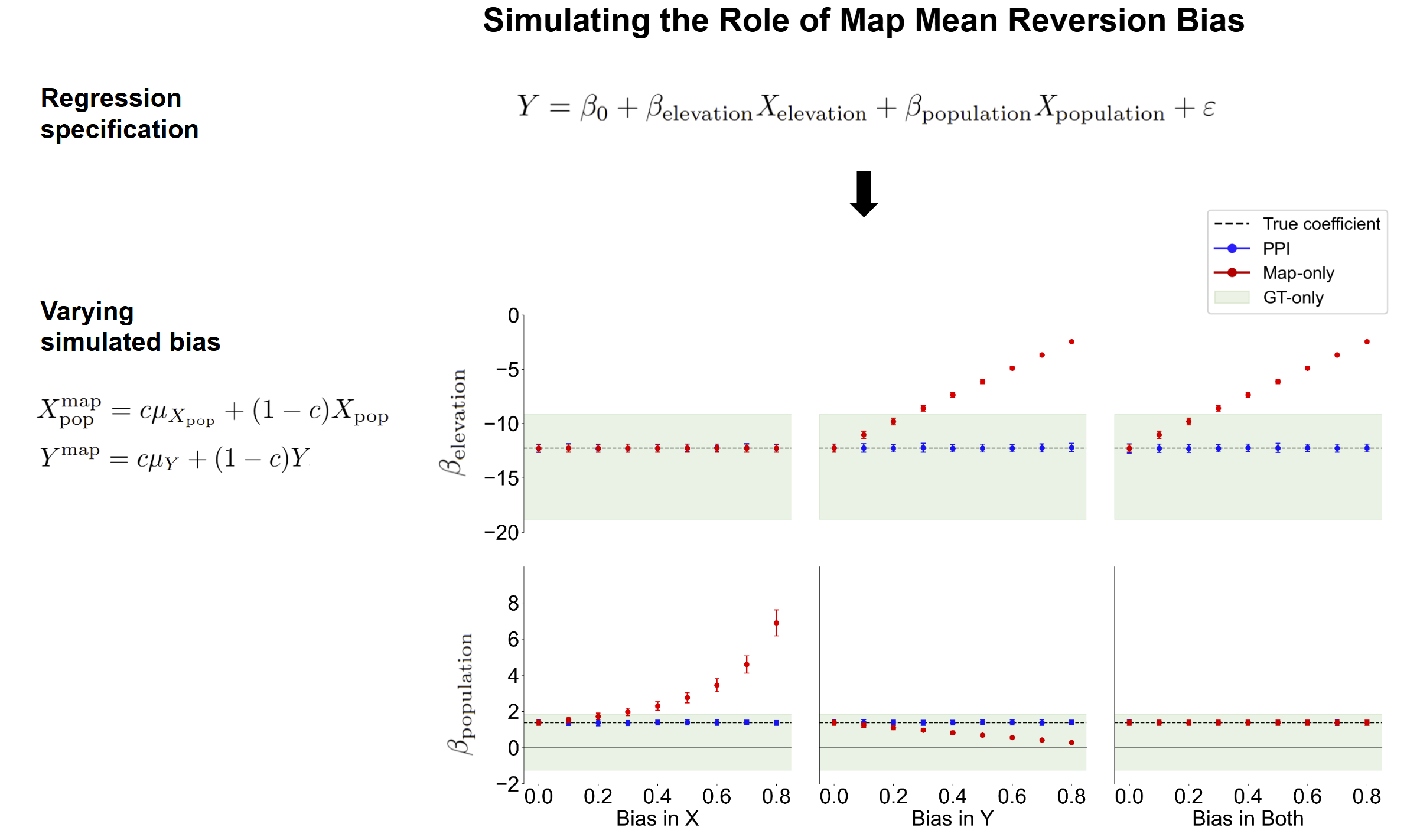}

\caption{\textbf{Simulation estimating forest cover linear regression coefficient 95\% confidence intervals under different levels of map mean reversion bias.} PPI is unbiased and produces much narrower confidence intervals than the GT-only method. In each experiment, the map-only coefficient estimate for at least one covariate becomes increasingly biased as map bias increases.}\label{simulation-bias-mean-reversion}
\end{figure}

The coefficient point estimates and 95\% confidence intervals at each bias level are shown in Figure \ref{simulation-bias-mean-reversion}. We compare results for the map-only, GT-only, and PPI estimators. (For all experiments in this section, we use $B=200$ bootstrap iterations to construct PPI confidence intervals.)

Similar to Section \ref{sec:simulations}, the GT-only confidence interval stays constant regardless of the bias level, and it is wider than the PPI and map-only intervals due to the small number of ground truth data points ($n=500$). 

At all bias levels across all three experiments (error in $X_{\text{pop}}$, $Y$, or both), the PPI confidence intervals contain the true coefficients for both covariates. This is expected, as PPI uses the small set of ground truth data points to correct for bias in the map product, and the PPI confidence intervals are guaranteed to contain the true coefficient with 95\% probability. 

At all bias levels, the PPI confidence intervals are significantly narrower than the GT-only intervals. This is because the biased map is a simple linear transformation of the true variable values and thus provides useful information for estimating the regression coefficients. As discussed in Section \ref{sec:simulations}, PPI allows us to conclude that there is a positive association between forest cover and population, while the GT-only interval for $\beta_{\text{pop}}$ contains zero and does not detect this significant association. 

In all three experiments, as map bias increases, the map-only estimates become biased for at least one of the coefficients. The map-only confidence intervals are very narrow because the number of map product data points is large ($N = 67968$), so they fail to contain the true coefficients as the estimates become biased. As mean reversion bias in $X_{\text{pop}}$ increases, the map underestimates high population values and overestimates low population values. As a result, the map-only estimates for $\beta_{\text{pop}}$ increase in magnitude and overestimate the true coefficient, while map-only confidence intervals for $\beta_{\text{elevation}}$ contain the true coefficient (since we do not add bias to the elevation map). As mean reversion bias in $Y$ (forest cover) increases, the map underestimates high forest cover values and overestimates low forest cover values. As a result, the map-only coefficient estimates for both covariates decrease in magnitude and attenuate toward zero. When there is mean reversion at level $c$ in both forest cover and population, the effects of the two biases cancel out to produce map-only confidence intervals for $\beta_{\text{pop}}$ that contain the true coefficient. However, the map-only estimates of $\beta_{\text{elevation}}$ still decrease in magnitude and attenuate toward zero. 

\clearpage
\section{Regression estimator for mean or area estimation}\label{appendix:regression-estimator}
In this appendix, we describe the regression estimator for mean estimation \cite{cochran1977sampling, gallego2004remote}, which we will refer to as the \textbf{regression mean estimator} and how this estimator relates to the PPI estimator. (Note that area estimation can be formulated as estimating the mean of a binary variable representing a land use/land cover class.)

Suppose we wish to estimate the mean $\theta^*=\mathbb{E}[Y]$ of a variable $Y$ that is only observed on a small simple random sample. Suppose we also observe a vector of one or more auxiliary variable(s) $X$ over the entire population. Let $\bar{y}$ be the sample mean of $Y$, let $\bar{x}$ be the vector of sample means of $X$ from the same sample where $Y$ was observed, and let $\bar{X}$ be the vector of population means of $X$. Let $b$ be the vector of linear regression coefficients obtained by regressing $Y$ against $X$ over the sample where $Y$ was observed. The regression estimator of the mean of $Y$ is
\begin{align}\label{formula:regression-mean-estimator}
    \hat{\theta}_{\text{reg}} = \bar{y} + b^T(\bar{X}-\bar{x}).
\end{align}
Note that the vector of auxiliary variables $X$ can contain covariate(s) that are different from $Y$, or it can be a proxy for $Y$ itself. In the latter case, $X = \hat{Y}$ are taken from a remotely sensed map with predictions of $Y$, and the regression mean estimator performs mean estimation by combining a small random sample of ground truth data points with map data for the same variable.

Although this estimator involves a regression, the estimator's purpose is to estimate the mean of a variable rather than estimate regression coefficients relating different variables. We also note that the linear regression coefficient vector $b$ can be the same as the GT-only linear regression coefficient estimator from Section \ref{subsection:linear-regression} of the main text, since only the ground truth data points from the small sample are used. By contrast, PPI uses both the ground truth calibration sample and map product data to estimate regression coefficients. 

However, there are some notable conceptual similarities between the regression mean estimator and the PPI regression coefficient estimator described in Section \ref{subsection:linear-regression}. The PPI estimator for a vector of regression coefficients $\beta$ can be written as
\begin{align*}
    \hat{\beta}_{\text{PPI}} &= \hat{\Omega} \hat{\gamma}_{\text{map}} + (\hat{\beta}_{\text{calib}} - \hat{\Omega} \hat{\gamma}_{\text{calib}}) \\
    &= \hat{\beta}_{\text{calib}} + \hat{\Omega}(\hat{\gamma}_{\text{map}} - \hat{\gamma}_{\text{calib}})
\end{align*}
where 
\begin{itemize}
    \item $\hat{\beta}_{\text{calib}}$ is the regression coefficient estimator using the ground truth values from the calibration sample,
    \item $\hat{\gamma}_{\text{map}}$ is the regression coefficient estimator using all map product values,
    \item $\hat{\gamma}_{\text{calib}}$ is the regression coefficient estimator using the map product values from the calibration sample, and
    \item $\hat{\Omega}$ is a tuning matrix that minimizes the variance of each component of $\hat{\beta}_{\text{PPI}}$.
\end{itemize}
Thus, the PPI estimator for the $i$th component of $\beta$ is
\begin{align*}
    \hat{\beta}_{\text{PPI}}^{(i)} = \hat{\beta}_{\text{calib}}^{(i)} + \hat{\Omega}_{i\cdot}^T(\hat{\gamma}_{\text{map}} - \hat{\gamma}_{\text{calib}})
\end{align*}
where $\hat{\beta}_{\text{calib}}^{(i)}$ denotes the $i$th component of $\hat{\beta}_{\text{calib}}$, and $\hat{\Omega}_{i\cdot} \in \mathbb{R}^d$ denotes the $i$th row of $\hat{\Omega}$.

Comparing this to the regression mean estimator (Equation \ref{formula:regression-mean-estimator}), we can view $\hat{\beta}_{\text{calib}}^{(i)}$, $\hat{\gamma}_{\text{map}}$, $\hat{\gamma}_{\text{calib}}$, and $\hat{\Omega}_{i\cdot}$ as the PPI analogues of $\bar{y}$, $\bar{X}$, $\bar{x}$, and $b$ respectively. In particular, both $\hat{\beta}_{\text{calib}}^{(i)}$ and $\bar{y}$ are scalars computed using a small sample of ground truth values; both $\hat{\gamma}_{\text{map}}$ and $\bar{X}$ are vectors computed using a large dataset of ``proxy" values (i.e., map product values or $X$ values); both $\hat{\gamma}_{\text{calib}}$ and $\bar{x}$ are vectors computed using the proxy values from the small sample; and both $\hat{\Omega}_{i\cdot}$ and $b$ are vectors designed to reduce the variance of the estimators. The key difference is that $\hat{\beta}_{\text{calib}}^{(i)}$, $\hat{\gamma}_{\text{map}}$, and $\hat{\gamma}_{\text{calib}}$ are regression coefficient estimators while $\bar{y}$, $\bar{X}$, and $\bar{x}$ are means. 

\newpage
\section{Applying PPI to mean and area estimation}
In this appendix, we demonstrate how to apply PPI to estimating the mean of a remotely sensed variable or the area of a land cover class. In Section \ref{appendix:ppi-mean-estimation}, we describe the PPI estimator for mean estimation. In Section \ref{appendix:area-estimation}, we apply PPI to two area estimation tasks (Iowa maize area and Brazilian Amazon deforestation area), and compare results with stratified and post-stratified estimators. In Section \ref{appendix:ppi-post-stratified-equivalent}, we mathematically show that PPI and the post-stratified estimator produce similar confidence intervals. 

\subsection{PPI for mean estimation}\label{appendix:ppi-mean-estimation}

In this section, we describe how to use PPI to estimate the mean value of a remotely sensed variable. We describe ground truth-only, map-only, and PPI estimators. (Note that \textbf{area estimation} can be formulated as estimating the mean of a binary variable representing a land use/land cover class. See Section \ref{appendix:area-estimation} for examples applying PPI for area estimation on remote sensing datasets.)

\paragraph{Ground truth-only estimator.}
First, \textbf{ground truth-only (GT-only)} uncertainty quantification uses ground truth observations, selected via simple random sampling, to estimate the quantity of interest. For example, suppose there is a response variable $Y$ that can be observed over a given area, and we wish to generate a 95\% confidence interval for its population mean $\theta^{\star} = E[Y]$. If we have a simple random sample of $n$ ground truth observations $Y_1, Y_2, \dots Y_n$, we can compute the sample mean 
\begin{align*}
    \hat{\theta} = \frac{1}{n} \sum_{i=1}^n Y_i, \quad \text{which has standard deviation } \quad \sigma_{\hat{\theta}}
    = \sqrt{\frac{1}{n} \Var(Y_i)},
\end{align*}
and estimate the standard deviation with
\begin{align*}
    \hat{\sigma}_{\hat{\theta}}  
 = \frac{1}{n} \sqrt{\sum_{i=1}^n (Y_i - \hat{\theta})^2}
\end{align*}
to produce the GT-only 95\% confidence interval
\begin{align*}
    C = (\hat{\theta} - 1.96 \cdot \hat{\sigma}_{\hat{\theta}}, \ \hat{\theta} + 1.96 \cdot \hat{\sigma}_{\hat{\theta}}).
\end{align*} If the size $n$ of the ground truth dataset is small, which is often the case in environmental applications, the GT-only confidence interval will be wide.

\paragraph{Map-only estimator: Use only the remote sensing map.}
If we have a remote sensing data product with a large number $N$ of predictions $\hat{Y}_1, \hat{Y}_2, \dots \hat{Y}_N$ over the area of interest, we could treat these predictions like ground truth data points and use the the formula above with $\hat{Y}$ in place of $Y$. This would produce a much narrower confidence interval, since we would use $N \gg n$ in the denominator of the standard error. However, this \textbf{map-only estimator} may be biased, and the confidence interval is not guaranteed to have a 95\% coverage rate of the true parameter of interest. This is because the map-only estimator does not account for error or uncertainty in the model used to generate the map product predictions.  

\paragraph{Prediction-Powered Inference: Map + ground truth.}
PPI \cite{angelopoulos2023prediction} evaluates the accuracy of the map product by comparing the ground truth labels $Y_1, Y_2, \dots, Y_n$ to the map product predictions $\hat{Y}_1, \hat{Y}_2, \dots, \hat{Y}_n$ at the corresponding ground truth data point locations. This \textbf{calibration set} of $n$ data points is a simple or stratified random sample over the region of interest, and it must be separate from the training dataset used to train the machine learning model. PPI uses the calibration set to correct for bias in the map-only estimator computed from the $N$ unlabeled map data points $\hat{Y}'_1, \hat{Y}'_2, \dots \hat{Y}'_N$.

For a concrete example, suppose we wish to estimate $\theta^{\star} = E[Y]$ as before. In the special case of estimating $\theta^{\star} = E[Y]$, the PPI point estimator is  
\begin{align*}
    \hat{\theta}_{\text{PPI}} = \underbrace{\frac{1}{N}\sum_{i=1}^N \hat{Y}_i'}_{\text{map-only estimator}} - \underbrace{\frac{1}{n} \sum_{i=1}^n (\hat{Y}_i - Y_i)}_{\text{bias correction}},
\end{align*} which coincides precisely with an estimator previously studied in the remote sensing literature (see Equation 4c of \cite{MCROBERTS2022113168}).
Assuming the map sample and calibration sample are independent, the standard deviation of this estimator is
\begin{align*}
    \sigma_{\text{PPI}} &= \sqrt{\frac{1}{N} \Var(\hat{Y}_i') + \frac{1}{n} \Var(\hat{Y}_i - Y_i)} \\ 
    &\approx \sqrt{\frac{1}{n} \Var(\hat{Y}_i - Y_i)},
\end{align*}
where the approximation holds when $N$ is much larger than $n$. Letting $\hat{\sigma}_{\text{PPI}}$ be an estimate of $\sigma_{\text{PPI}}$ based on plugging sample variance estimates into the formula for $\sigma_{\text{PPI}}$, the PPI 95\% confidence interval is given by
\begin{align*}
    C_{\text{PPI}} = (\hat{\theta}_{\text{PPI}} - 1.96 \cdot \hat{\sigma}_{\text{PPI}}, \  \hat{\theta}_{\text{PPI}} + 1.96 \cdot \hat{\sigma}_{\text{PPI}}).
\end{align*}

The first term in $\sigma_{\text{PPI}}$ is the variance corresponding to using only map predictions to estimate $\theta$, i.e. $\frac{1}{N} \Var(\hat{Y}_i')$. The second term is the variance from the bias correction, $\frac{1}{n} \Var(\hat{Y}_i - Y_i)$. As the number of map product data points $N$ grows large, $\sigma_{\text{PPI}}$ approaches $\sqrt{\frac{1}{n} \Var(\hat{Y}_i - Y_i)}$. This will be a realistic approximation in many cases, since remote sensing maps at a global, national, or province level can contain millions or even billions of pixels, whereas the number of ground truth data points $n$ is commonly in the hundreds or thousands. The greater the accuracy of the map product, the smaller $\Var(\hat{Y}_i - Y_i)$ will be and the narrower the PPI confidence interval will be. Note that the PPI confidence interval is narrower than the GT-only confidence interval if $\Var(\hat{Y}_i - Y_i) < \Var(Y_i)$. 


Like PPI, the post-stratified area estimator \cite{card1982using, stehman2013estimating, olofsson2013making} is unbiased, combining a map product with a small amount of ground truth data to make a precise estimate. The post-stratified estimator and PPI produce similar outputs for area estimation; for more detailed comparisons of the two methods, see Sections \ref{appendix:area-estimation} and \ref{appendix:ppi-post-stratified-equivalent}. However, the post-stratified estimator only applies to area estimation, not regression coefficient estimation, while PPI can be applied to both (Figure \ref{summary} in the main text). 

Similarly, the model-assisted regression estimator from \cite{MCROBERTS2022113168} coincides precisely with the PPI estimator in the context of mean estimation, but does not apply to the setting of regression coefficient estimation.

\subsection{Area estimation experiments}\label{appendix:area-estimation}
In this section, we apply PPI and other methods to \textbf{area estimation} tasks with both ground truth and remotely sensed data.

We first describe two well-known methods for area estimation: (1) stratified estimator with a stratified sampling design and (2) post-stratified estimator with a simple random sampling design. These methods were introduced in \cite{cochran1977sampling} and are commonly used in the remote sensing community \cite{olofsson2013making, olofsson2014good}. PPI for area estimation (described in Section \ref{appendix:ppi-mean-estimation}) produces similar confidence intervals as the post-stratified estimator; for a mathematical comparison, see Section \ref{appendix:ppi-post-stratified-equivalent}. 

Then, we apply PPI, the stratified estimator, and the post-stratified estimator to two area estimation use cases (Iowa maize crop area and Brazilian Amazon deforestation area). We also compare results with the GT-only and map-only mean estimation methods described in Section \ref{appendix:ppi-mean-estimation}. 

\subsubsection{Stratified estimator}\label{appendix:stratified-estimator}
\textbf{Stratified random sampling} uses remotely sensed  maps to determine a sampling strategy for ground truth data points that reduces the variance of estimators. The \textbf{stratified estimator} is commonly used in the remote sensing community.

We first use a map product to partition the area of interest into $K$ strata (sub-areas) such that the response variable is predicted to have low variance within each stratum. Then we take a simple random sample of $n_k$ ground truth data points from each stratum $1 \leq k \leq K$. We can choose $n_k$ to be larger for strata predicted to have higher uncertainty to decrease the standard error within the strata and hence make the overall standard error smaller with the same total number of sample units.

Suppose we observe the response variable $Y$ over a given area, and we wish to estimate $\theta^{\star} = E[Y]$ using a stratified random sample. For each stratum $k$, we compute the stratum sample mean $\hat{\theta}_k$ and stratum standard error $\hat{\sigma}_k$ using the same formulae as the GT-only method from Appendix \ref{appendix:ppi-mean-estimation}. Suppose that in the map product, each stratum $k$ takes up proportion $0 < A_k < 1$ of the total area (with $\sum_{k=1}^{K} A_k=1$). Then, we compute the stratified mean 
\begin{align*}
    \hat{\theta}_{\text{strat}} = \sum_{k=1}^K A_k \hat{\theta}_k
\end{align*}
and stratified standard error
\begin{align*}
    \hat{\sigma}_{\text{strat}} = \sqrt{\sum_{k=1}^K A_k^2 \cdot \hat{\sigma}_k^2} 
\end{align*}
to produce the stratified 95\% confidence interval
\begin{align*}
    C_{\text{strat}} = (\hat{\theta}_{\text{strat}} - 1.96 \cdot \hat{\sigma}_{\text{strat}}, \ \hat{\theta}_{\text{strat}} + 1.96 \cdot \hat{\sigma}_{\text{strat}}).
\end{align*}

The advantage of stratified estimation with an accurate map and well-chosen sampling strategy is that the variance within each stratum will be small, which results in a narrower overall confidence interval compared to the GT-only method. If the map predictions are inaccurate, then the ground truth data points likely have higher variance within each map stratum, resulting in a wider confidence interval. Thus the uncertainty quantification incorporates the accuracy of the map product predictions.

\begin{figure}[t]
\centering
\includegraphics[width=0.35\textwidth]{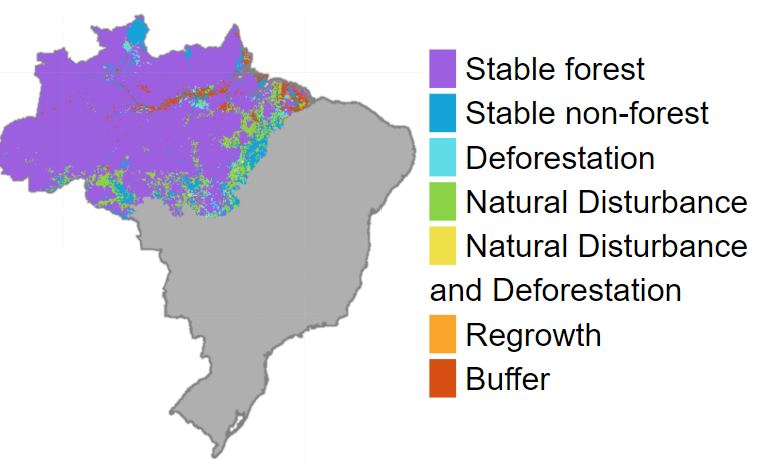}
\caption{\textbf{Stratified estimation is a commonly used approach with remote sensing maps}. This example shows a map of the Brazilian Amazon used for stratification. The strata are designed so that data points within each stratum have low variance in deforestation status, which reduces the uncertainty when estimating area deforested. Adapted from \cite{bullock2020satellite}.}\label{deforestation-strata}
\end{figure}

For example, in our Brazilian Amazon deforestation area estimation use case, we use a remotely sensed stratification map created by \cite{bullock2020satellite} using Landsat surface reflectance imagery at 30 m resolution. The map (Figure \ref{deforestation-strata}) partitions the Amazon into seven strata related to forest change from 1995-2017: stable forest, stable non-forest, deforestation, natural disturbance, natural disturbance and deforestation, regrowth, and buffer. The strata are designed so that data points within each stratum have low variance in deforestation status (for instance, most data points in the ``stable forest" and ``stable non-forest" strata do not experience deforestation in the time period of interest). As a result, our stratified confidence interval for deforestation area is narrower than the GT-only confidence interval. 

PPI can be used with either simple random sampling or stratified random sampling of the ground truth data points~\cite{fisch2024stratified}. \textbf{Stratified (weighted) PPI} can be applied as follows for mean estimation. Each ground truth value $Y_i$ and labeled prediction $\hat{Y}_i$ is multiplied by weight $w_i = A_k \frac{n}{n_k}$ where $k$ is the stratum where data point $i$ is located, $A_k$ is the area proportion of stratum $k$, $n_k$ is the number of ground truth sample units in stratum $k$, and $n = \sum_{i=1}^K n_k$. (Note that the weights are designed such that the sum of weights of all $n_k$ ground truth data points in stratum $k$ is $A_k n$, and the sum of weights of all $n$ ground truth data points is $\sum_{i=1}^K A_k n = n$. If the data points were selected via simple random sampling, this would be equivalent to only having a single ``stratum," and each data point would have weight 1.) Similarly, each unlabeled prediction $\hat{Y}'_i$ in stratum $k$ is multiplied by weight $w_i = A_k \frac{N}{N_k}$ where $N_k$ is the number of map product data points in stratum $k$ and $N = \sum_{i=1}^K N_k$. Then the same PPI formulae as in Section \ref{appendix:ppi-mean-estimation} can be used. 

\subsubsection{Post-stratified area estimator} 
The \textbf{post-stratified area estimator}, commonly used in the remote sensing community, utilizes the confusion matrix between ground truth and map product land cover labels to estimate the area of a land cover class \cite{card1982using, stehman2013estimating, olofsson2013making}. Like PPI, the post-stratified area estimator computes confidence intervals by using ground truth data points to correct for bias in the map-only estimator. This can be viewed as a correction that uses the map errors relative to the ground truth data points to modify the map-only area estimates. We show mathematically in Section \ref{appendix:ppi-post-stratified-equivalent} that PPI and the post-stratified area estimator result in similar confidence intervals as the number of map product data points $N$ grows large.
However, unlike PPI, the post-stratified area estimator only applies to area estimation but not regression coefficient estimation.

Turning to the method, suppose we have a map product with $K$ land cover classes, as well as $n$ randomly sampled ground truth data points. The post-stratified area estimator constructs the $K \times K$ misclassification confusion matrix between ground truth data points and map product predictions at the ground truth data point locations. For each $1 \leq i \leq K$ and $1 \leq j \leq K$, let $n_{ij}$ denote the number of data points that are in map class $i$ and ground truth class $j$. Also let $n_{i\cdot}=\sum_{j=1}^K n_{ij}$ denote the total number of ground truth sample units that have map class $i$. For each class $1 \leq i \leq K$, let $0 < A_i < 1$ denote the proportion of the total map area taken up by the map class. 

Then, motivated by the formula $\mathbb{P}(Y=j)=\sum_{i=1}^K \mathbb{P} (\hat{Y}=i) \cdot \mathbb{P} (Y=j \ | \ \hat{Y}=i)$, the post-stratified area estimator for the proportion of area of class $j$ is
\begin{align*}
    \hat{\theta}_{j, \text{post}} = \sum_{i=1}^K A_i \frac{n_{ij}}{n_{i\cdot}}
\end{align*}
with corresponding standard error
\begin{align*}
    \hat{\sigma}_{j, \text{post}} = \sqrt{\sum_{i=1}^K A_i^2 
    \frac{\frac{n_{ij}}{n_{i\cdot}} (1-\frac{n_{ij}}{n_{i\cdot}})}{n_{i\cdot}}}.
\end{align*}
The 95\% confidence interval for the proportion of area of class $j$ is
\begin{align*}
    C_{j, \text{post}} = (\hat{\theta}_{j, \text{post}} - 1.96 \cdot \hat{\sigma}_{j, \text{post}}, \ \hat{\theta}_{j, \text{post}} + 1.96 \cdot \hat{\sigma}_{j, \text{post}}).
\end{align*}

Area estimation is a special case of mean estimation in which we estimate the mean of a binary variable that is $1$ for all data points in class $j$ and $0$ for all data points outside of class $j$. Thus, we can also perform area estimation using methods such as PPI and stratified estimation.

If the $n$ ground truth data points are obtained by stratified random sampling using the $K$ map land cover classes as the strata, the post-stratified confidence interval computed using this data is identical to the stratified confidence interval \cite{olofsson2013making}. However, unlike the stratified estimator, the post-stratified estimator can be applied even if the ground truth data points are obtained using simple random sampling.  

\subsubsection{Area estimation examples}

In this section, we demonstrate uncertainty quantification for two area estimation use cases. First, we simulate estimating maize area in Iowa in 2022, artificially degrading the map product quality relative to simulated ``ground truth" data points, in order to observe the effect of map product noise and bias on the area estimate under different methods. We will see through this simulation the importance of bias correction before using maps to estimate total area. We then use real ground truth data and real map products developed by researchers to estimate 
change in tree cover in the Brazilian Amazon in 2000--2015. We compare the GT-only, map-only (pixel-counting), stratified, post-stratified, PPI, and stratified PPI estimators.

In each use case, we use binary data where each data point has value 1 if it belongs to the land cover class of interest (e.g., cropland or deforested land), and value 0 otherwise. Then, estimating the area fraction of the land cover class is equivalent to estimating the mean of all data points in the region. 

We use the \textsc{ppi-python} software package to implement the unweighted and weighted versions of PPI for our area estimation use cases. The software includes a tuning parameter $\lambda$ that we set using the map product data points and ground-truth data points. The confidence intervals are guaranteed to have the appropriate (e.g., 95\%) coverage of the true area fraction even with this data-driven tuning~\cite{angelopoulos2023ppi++}.

Data and code for these use cases are available at \url{https://github.com/Earth-Intelligence-Lab/uncertainty-quantification}.

\paragraph{Simulation: Maize area in Iowa}
\paragraph{Problem Formulation}
We simulate estimating the fraction of maize area in Iowa in 2022. To probe the effect of map product noise and bias on different methods, we artificially degrade the map product quality relative to simulated ``ground truth" data points, as described next. 

\begin{figure}[b]
\centering
\includegraphics[width=0.4\textwidth]{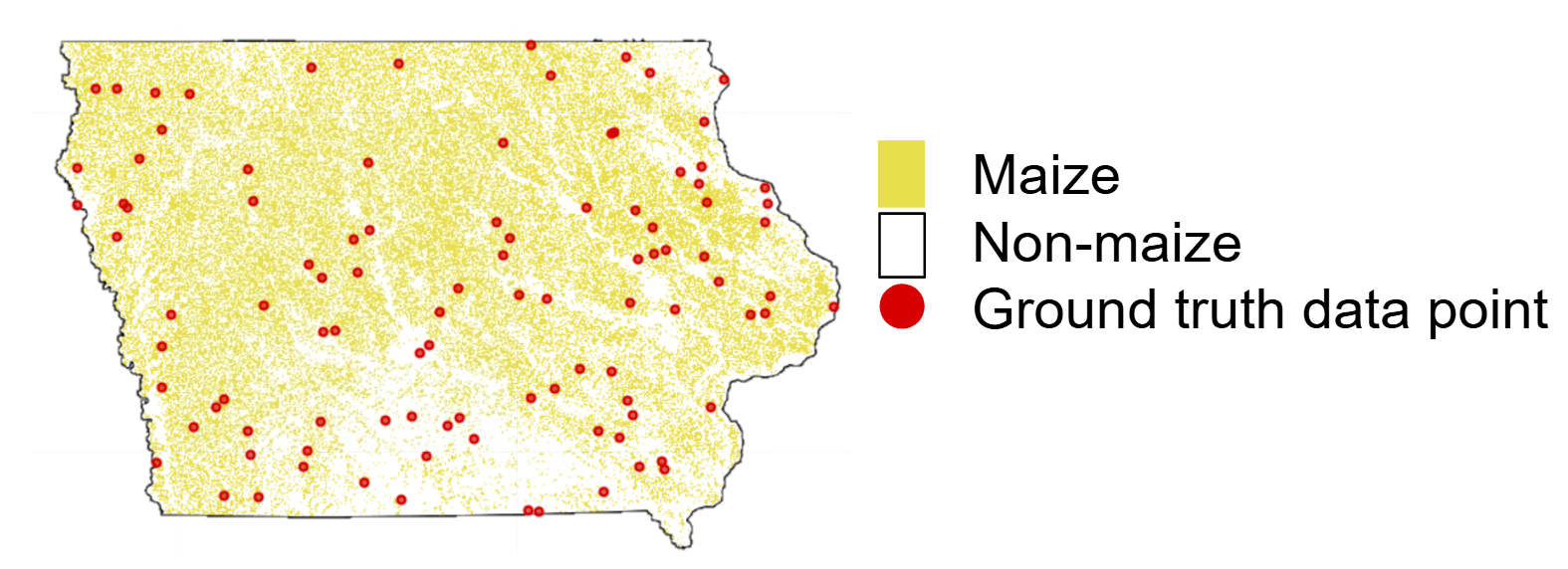}
\caption{\textbf{Iowa maize area estimation datasets.} The Iowa maize binary map product is derived from the USDA's Cropland Data Layer, and a simple random sample of $n=100$ ``ground truth data points" are shown.}\label{corn-map}
\end{figure}

For this simulation, we use the USDA's Cropland Data Layer (CDL) \cite{nass2024usda} as the ground truth for crop types in Iowa. CDL is a crop type land cover map created using satellite imagery and agricultural ground truth data in the continental United States. It has 94.2\% producer's accuracy and 97.2\% user's accuracy for the maize land cover class in Iowa. We take a simple random sample of $N=100000$ locations from a binary maize/non-maize map derived from CDL (Figure \ref{corn-map}). The mean of the 100000 map product data points is 0.35, which we will consider the ``true" fraction of maize for the simulation.  

In each trial, we take a simple random sample of $n=100$ ``ground truth" data points from the $N=100000$ map product data points. Initially, we set the map product and ground truth equal to each other at the 100 ground truth locations. We then experiment with degrading the map product quality as follows. We systematically add noise or bias to the map product data points while keeping the ground truth data points constant, and then compare our maize fraction estimates across the map-only (pixel-counting), GT-only, and PPI estimators. We average results over 100 trials. 

\paragraph{Adding unbiased noise.} We define adding ``unbiased noise" at level $p$ to the map product as resampling each of the 100000 map product data points with probability $p$ from the Bernoulli(0.35) distribution. In other words, about $p$ fraction of the data points are randomly labeled, while the remaining $1-p$ data points are unchanged. Each resampled data point is labeled as maize with probability 0.35 and non-maize with probability 0.65. We experiment with noise probabilities $p$ from 0 to 1 at increments of 0.1.

\paragraph{Adding bias.} We also consider adding bias to the map product. Here, we define adding bias $p$ to the map product as resampling each of the 100000 map product data points with probability 0.4 from the Bernoulli($p$) distribution. In other words, about 40\% of the data points are randomly re-labeled, while the remaining 60\% data points are unchanged. (After the re-labeling, more than 60\% of the data points will have the same label as the original map.) Each resampled data point is labeled as maize with probability $p$ and non-maize with probability $1-p$. We experiment with bias values $p$ from 0 to 1 at increments of 0.1. When $p = 0.35$, this does not add bias, but as $p$ ranges farther from $0.35$, the amount of bias increases.

\paragraph{Results}

\begin{figure}[t]
\centering
\includegraphics[width=0.8\textwidth]{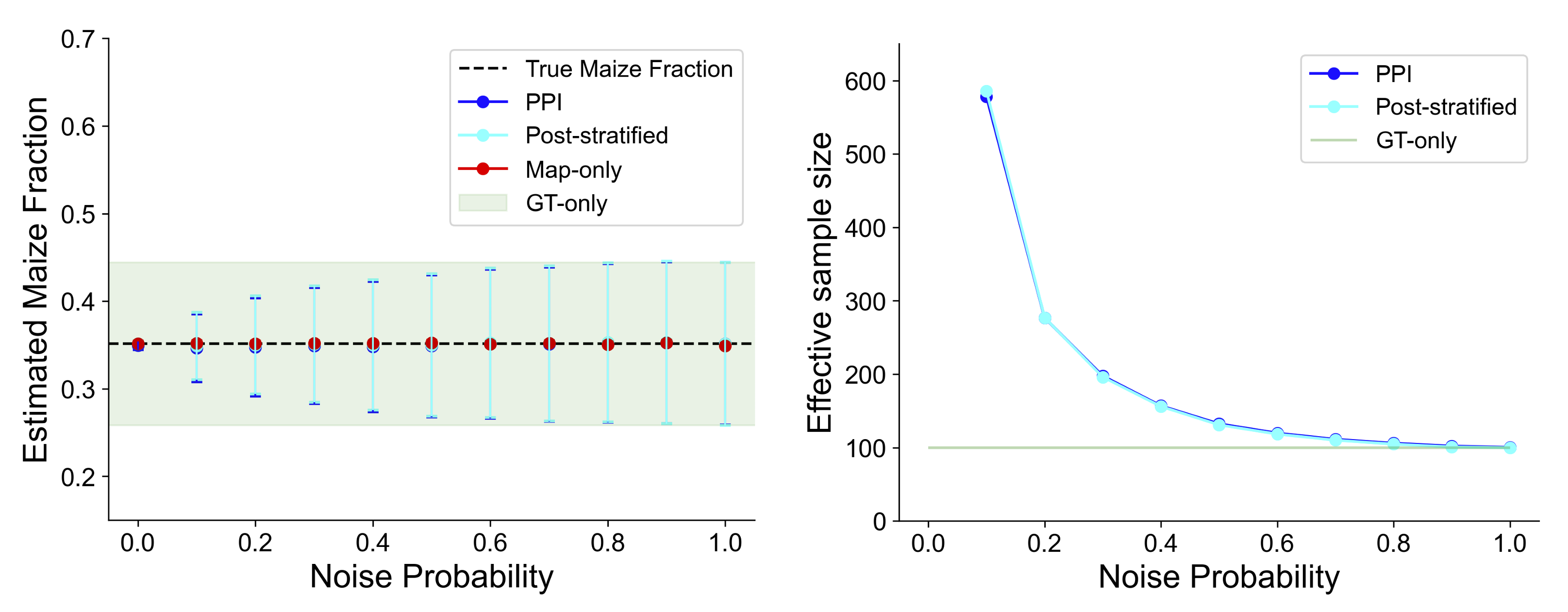}
\caption{\textbf{Simulation estimating maize fraction under different levels of map noise} using $N=100000$ map product data points and $n=100$ ``ground truth" data points. Post-stratified and PPI results are overlapping. Left: Maize fraction 95\% confidence intervals at different noise levels. Right: Maize fraction effective sample size at different noise levels. (At noise level $p$, about $p$ fraction of the data points are resampled, while the remaining $1-p$ data points are unchanged. Each resampled data point is labeled as maize with probability 0.35 (the true maize fraction) and non-maize with probability 0.65.)}\label{corn-noise}
\end{figure}

As we add increasing amounts of random noise to the map, all confidence intervals for maize area remain centered at the true maize fraction of 0.35 (Figure \ref{corn-noise}). This is what we expect, since our noise-adding process resamples from the Bernoulli(0.35) distribution, causing the overall fraction of maize to remain the same. The GT-only confidence interval stays constant because it only uses ground truth data points. The map-only confidence intervals are narrow because $N=100000$ is large, but they do not account for error in the map product. However, the map-only confidence intervals contain the true maize fraction because the map still has true maize fraction close to 0.35 regardless of noise level.

When the noise probability is low, the PPI confidence intervals are significantly narrower than the GT-only intervals. This is because the error of the map product ($\hat{Y} - Y$) is small at the ground truth locations, so the variance of this error is also small, resulting in a small PPI standard error. As noise increases, the variance of the error of the map product increases; the PPI confidence intervals grow wider, eventually converging to the GT-only confidence interval at noise probability $p=1$. (At $p=1$, the map is complete noise and therefore does not provide any information for estimating the area beyond the information contained in the ground truth sample.) Consequently, the PPI effective sample size decreases from $n_{\text{effective}} = 578$ when $p=0.1$ to $n_{\text{effective}} = 100$ when $p=1$ (which matches the number of actual ground truth data points $n=100$). Thus, the effectiveness of PPI is dependent on the quality of the map product. The post-stratified confidence intervals and effective sample sizes are nearly identical to those of PPI.

\begin{figure}[t]
\centering
\includegraphics[width=0.4\textwidth]{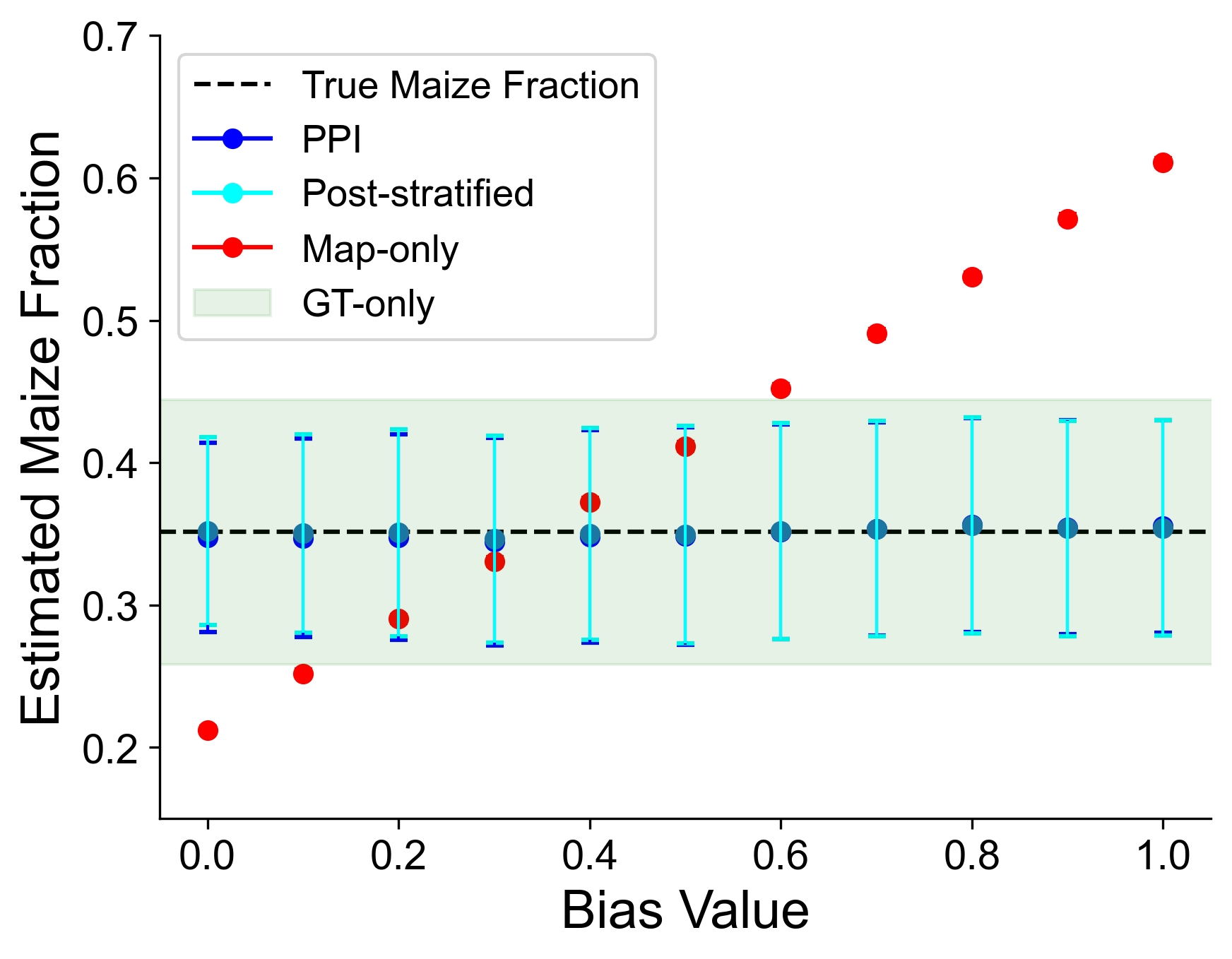}
\caption{\textbf{Simulation estimating maize fraction 95\% confidence intervals under different levels of map bias} using $N=100000$ map product data points and $n=100$ ``ground truth" data points. Post-stratified and PPI confidence intervals are overlapping. (Noise is fixed at 0.4 so about 40\% of the data points are resampled. At bias $p$, each resampled data point is labeled as maize with probability $p$ and non-maize with probability $1-p$.)}\label{corn-bias}
\end{figure}

The estimated maize fraction confidence intervals after adding bias to the map product are shown in Figure \ref{corn-bias}. Again, the GT-only confidence interval stays constant and remains centered at the true maize fraction because it only uses the ground truth data points. As bias increases, the fraction of map product data points labeled as ``maize" increases. As a result, the map-only estimates increase with the bias value, and generally do not remain centered at the true maize fraction of 0.35. The map-only confidence intervals are narrow because $N=100000$ is large, but they are unreliable because they do not correct for the bias in the map product. In contrast, the PPI estimates remain centered at the true maize fraction because PPI corrects for bias in the map product. The PPI confidence intervals are narrower than the GT-only intervals because the variance of the map product error $\hat{Y} - Y$ is small (since only about 40\% of the map product data points are resampled, regardless of bias level). Thus, PPI is unbiased and produces estimates with lower uncertainty than using only the ground truth data points. Again, the post-stratified confidence intervals and effective sample sizes are nearly identical to those of PPI, as expected.

\begin{figure}[t]
\centering
\includegraphics[width=0.8\textwidth]{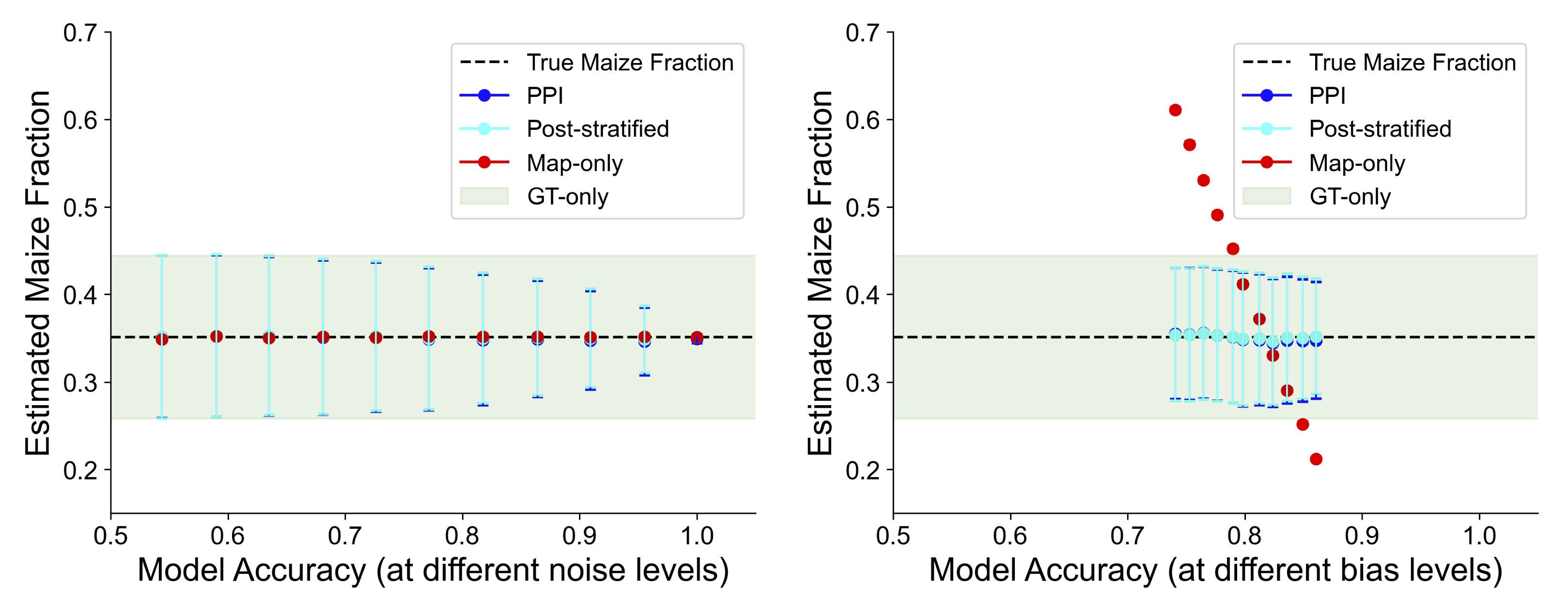}
\caption{\textbf{Simulation estimating maize fraction 95\% confidence intervals at different map product overall accuracy levels} obtained by varying noise (left) and bias (right), using $N=100000$ map product data points and $n=100$ ``ground truth" data points. Post-stratified and PPI confidence intervals are overlapping.}\label{corn-model-accuracy}
\end{figure}

We also plot the maize fraction confidence intervals against overall map accuracy for both the noise and bias experiments (Figure \ref{corn-model-accuracy}). As noise increases, map accuracy decreases, so low map accuracy is associated with wide PPI and post-stratified confidence intervals. Thus, the left panel of Figure \ref{corn-model-accuracy} resembles an inverted version of the left panel of Figure \ref{corn-noise}. 

As bias increases and the fraction of map product data points labeled as ``maize" increases, map accuracy decreases linearly. (To see this, note that if each resampled data point is labeled as maize with probability $p$, then an expected $0.35(p) + 0.65(1-p) = 0.65-0.3p$ fraction of the resampled data points will match the ground truth.) Thus, the map-only estimates of maize fraction decrease linearly as map accuracy increases. In the right panel of Figure \ref{corn-model-accuracy}, a map accuracy of 0.86 corresponds to a map-only maize fraction estimate of 0.21, significantly lower than the true fraction of 0.35. This shows that even a map with high map accuracy can yield biased estimates under the map-only (pixel-counting) estimator.

\paragraph{Deforestation area in Brazilian Amazon}
\paragraph{Problem Formulation}
\label{sec:deforestation-area}
Next, we move on to real data and estimate the fraction of deforestation in the Brazilian Amazon from 2000 to 2015, comparing confidence interval widths across methods. 

\begin{figure}[t]
\centering
\includegraphics[width=0.75\textwidth]{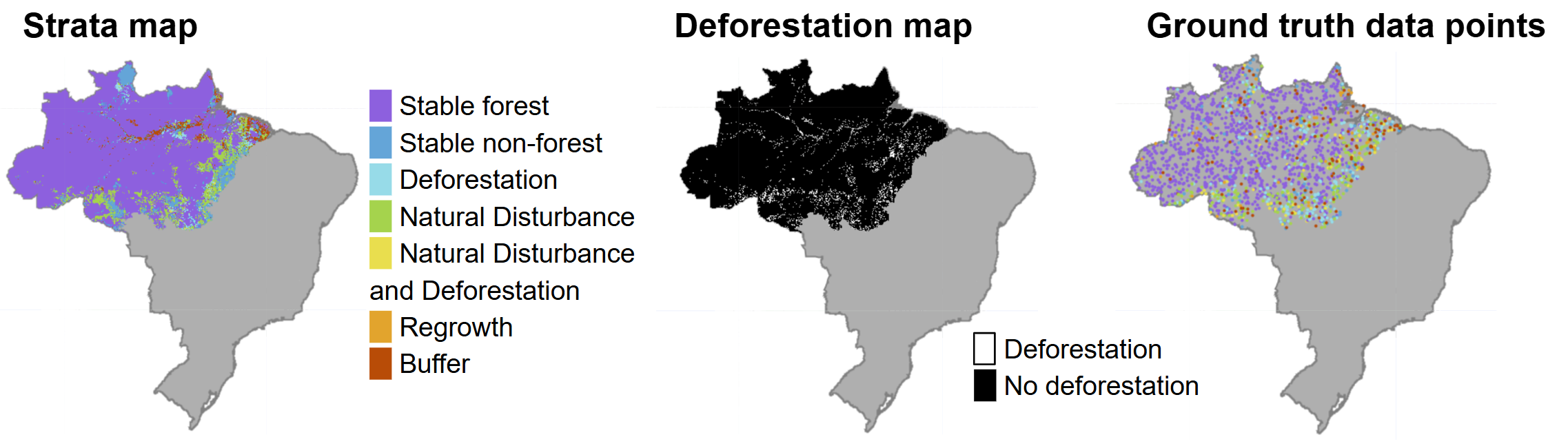}
\caption{\textbf{Brazilian Amazon deforestation area estimation datasets.} Left: Stratification map from \cite{bullock2020satellite}. Center: Binary map of deforestation from 2000-2015 derived from NASA Global Forest Cover Change. Right: $n=1386$ ground truth data points from \cite{bullock2020satellite}, colored according to the stratum from which they are sampled.}
\label{deforestation-map}
\end{figure}

We create a binary deforestation map product using all $N=4.4 \times 10^9$ pixels from the NASA Global Forest Cover Change (GFCC) map \cite{townshend2016global} (at 30 m resolution) in the study area (Figure \ref{deforestation-map}). The GFCC product is derived from Landsat 5 and Landsat 7 surface reflectance imagery, and includes maps for tree cover percentage in each pixel in five-year increments from 2000 to 2015. We consider a map pixel deforested if its canopy cover percentage satisfies $\text{canopy}_{2015} - \text{canopy}_{2000} \leq -25\%$ (Figure \ref{deforestation-binarization} in Appendix \ref{appendix:datasets}). 

We use a ground truth Amazon deforestation and forest disturbance dataset created by \cite{bullock2020satellite} using time-series analysis of Landsat imagery and high-resolution imagery from Google Earth between 1995 and 2017. The dataset records the years corresponding to deforestation, degradation, and natural disturbance events at each ground truth data point. We use this to construct a binary ground truth dataset restricted to Brazil, in which we consider a data point deforested if it experienced a deforestation event between 2000 and 2015. The original dataset is obtained by stratified random sampling, using a stratification map (also created by \cite{bullock2020satellite}) that contains 7 strata related to deforestation status. To generate a GT-only estimator baseline, we construct a ``simple random sample" ground truth dataset by downsampling the data points in each of the strata to match its area proportion. For a fair comparison, we also downsample the original stratified ground truth dataset to match the size of the ``simple random sample" dataset. This results in two ground truth datasets of $n=1386$ data points each, corresponding to simple random sampling and stratified random sampling. Both ground truth datasets are used with PPI to obtain confidence intervals.

\paragraph{Results}

\begin{figure}[t]
\centering
\includegraphics[width=0.35\textwidth]{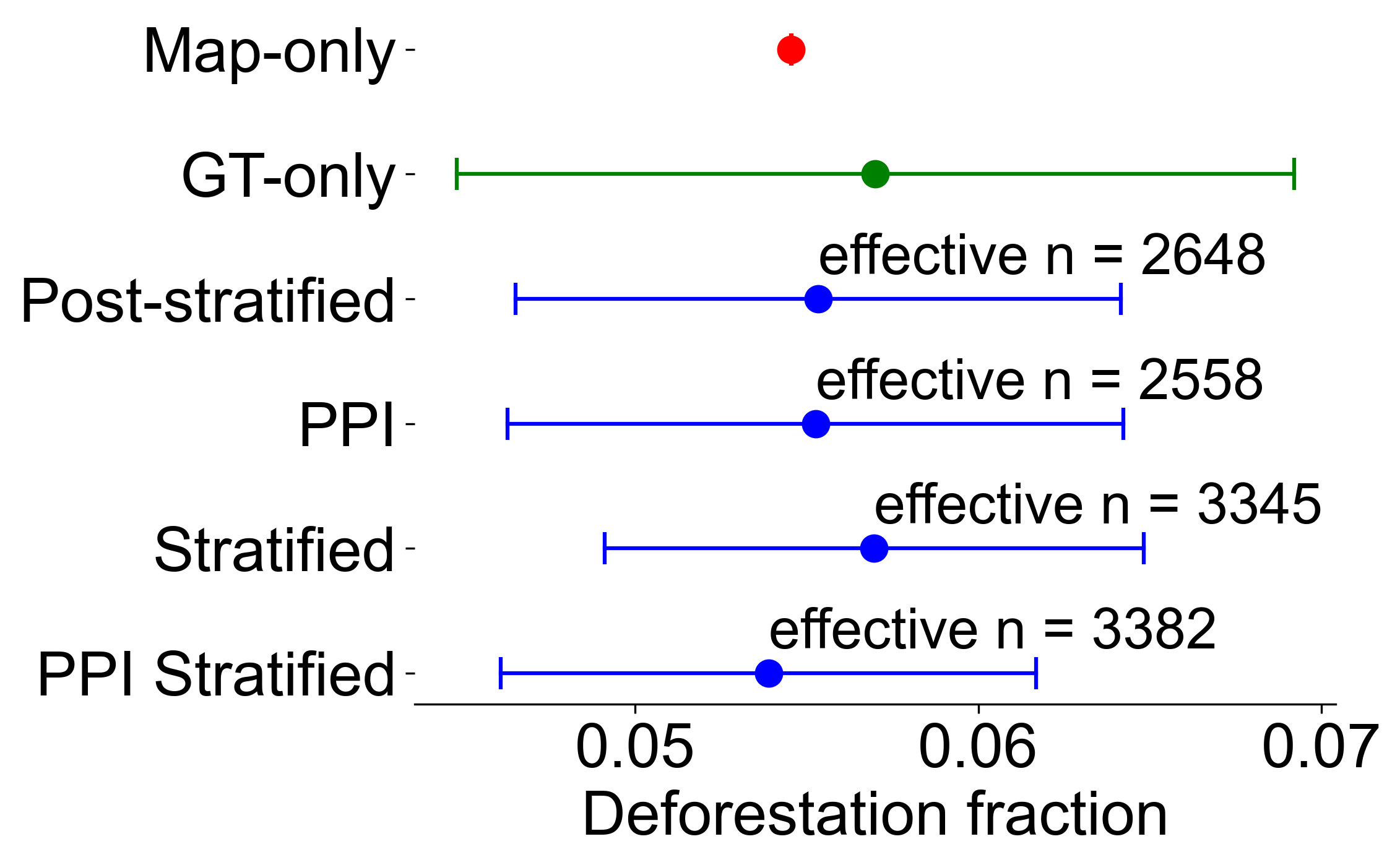}
\caption{\textbf{Brazilian Amazon deforestation 95\% confidence intervals} using $N=4.4 \times 10^9$ map product data points and $n=1386$ ground truth data points. The stratified, post-stratified, and PPI estimators all result in narrower confidence intervals for estimated deforestation than GT-only estimation, with the stratified estimator and stratified PPI having the best performances.}\label{deforestation-area}
\end{figure}

Each method results in a point estimate that between 5\% to 6\% of the Brazilian Amazon was deforested between 2000 and 2015. The stratified, post-stratified, and PPI estimators all result in narrower 95\% confidence intervals for estimated deforestation than GT-only estimation (Figure \ref{deforestation-area}). The post-stratified and PPI confidence intervals are fairly similar, with respective effective sample sizes that are $1.9$ and $1.8$ times the actual number of ground truth sample data points. The stratified and stratified PPI confidence intervals have the best performance, with effective sample sizes that outperform the GT-only method by a factor of about $2.4$. (Put differently, stratified PPI has effective sample size $3382$, so using the remotely sensed deforestation map added the equivalent of $3382-1386 = 1996$ ground truth sample data points.) Thus, the methods that use both map product and ground truth reduce uncertainty in our estimates, compared to using only ground truth. In particular, the stratified estimator is effective because the strata map is designed so that data points within each stratum have low variance in deforestation status.

In this case, the map-only (pixel-counting) estimate --- using only the map to estimate deforestation without any bias correction via ground truth labels --- falls within the other confidence intervals, suggesting that the NASA Global Forest Cover Change map has small bias for estimating deforestation area. However, we emphasize that this would not have been clear without the ground truth data, and ground truth data are still necessary to check the reliability of the map product for deforestation analysis. 

\subsubsection{Discussion}
\paragraph{Bias is the main issue for the map-only estimator.} Our experiments illustrated that a map-only area estimator from a map without bias correction can yield incorrect inferences, and we differentiated between the influence of noise versus bias. In our simulation estimating maize fraction, map-only estimates deviated from the true maize fraction of 0.35 as the bias of the crop type map increased. Furthermore, map accuracy is not necessarily a good proxy for bias; even at seemingly high map accuracies ($\text{accuracy} > 0.85$), the map-only approach estimated maize fraction at 0.21 (a 40\% underestimate).

\paragraph{Stratification vs. Post-stratification vs. PPI.} How much stratified, post-stratified, and PPI estimators reduce uncertainty relative to the GT-only estimator depends on the quality of the map product. In stratification, the strata map should be designed such that the variance of the quantity of interest is low within each stratum. In PPI (and the post-stratified method), the variance of the difference between the map product predictions $\hat{Y}$ and the ground truth values $Y$ should be low. In our maize fraction experiments, a highly accurate map ($\text{accuracy}=0.96$) was equivalent to having nearly $6\times$ as much ground truth data, whereas a low-accuracy map ($\text{accuracy}=0.73$) only increased the effective sample size by $1.2\times$.

Empirically, when estimating deforestation fraction in the Brazilian Amazon, we observed the narrowest confidence intervals when stratification was used alone or in conjunction with PPI. This is a result of well-designed strata based on a remote sensing map of forest disturbance categories. PPI and the post-stratified estimator yielded comparable interval widths to each other. This is expected, as the post-stratified estimator yields similar confidence intervals as PPI when $N$ is large (Section~\ref{appendix:ppi-post-stratified-equivalent}). 



An advantage of PPI compared to the stratified estimator is its flexibility. While the stratified estimator requires us to choose a map product before sampling, PPI can be applied to any map product and any randomly sampled ground truth dataset. Given the expensive nature of ground truth data collection, sampling without choosing a strata map and then using PPI to obtain confidence intervals may result in the best performance across a variety of downstream use cases. Or, if a stratified sample is already available from area estimation, it can be combined with PPI to estimate regression coefficients. 

\subsection{Mathematical comparison of PPI and post-stratified area estimator}\label{appendix:ppi-post-stratified-equivalent}
We show that when the number of map product data points $N$ is large, the PPI and post-stratified area estimator confidence intervals are similar. In particular, we show (1) that they have approximately the same width when the number of sample units is large and (2) they are centered on the same point. These two facts together provide a mathematical explanation for why PPI and the post-stratified estimator approximately agree in all of our experiments. 

We begin with notation. Estimating the area proportion of a land cover class in a given region is equivalent to estimating the mean of a binary variable $Y$, where $Y=1$ for data points in the class and $Y=0$ for data points outside the class. Suppose we have a simple random sample of $n$ ground truth data points $Y_1, Y_2, \dots Y_n$ and $N$ map product data points $\hat{Y}_1, \hat{Y}_2, \dots \hat{Y}_N$. For all $0 \leq i, j \leq 1$, let $n_{ij}$ denote the number of ground truth data points that are in both map class $i$ and ground truth class $j$. Let $n_{i \cdot} = n_{i0} + n_{i1}$ denote the total number of ground truth data points in map class $i$, and $n_{\cdot i} = n_{0i} + n_{1i}$ denote the total number of ground truth data points in ground truth class $i$. Furthermore, let $A_i$ denote the fraction of the map product that is in class $i$. We have $A_0 + A_1 = 1$. The map-only area estimator, which we will correct for bias, is $A_1$. We will use the approximations $A_0 \approx \frac{n_{0 \cdot}}{n}$ and $A_1 \approx \frac{n_{1 \cdot}}{n}$. This is a reasonable approximation since the $n$ ground truth locations are selected via simple random sampling. 

With this notation, we have the following,
\begin{align*}
    \sum_{i=1}^n Y_i &= n_{\cdot 1}\\
    \sum_{i=1}^n \hat{Y}_i &= n_{1 \cdot}\\
    \Var_n(Y) &= \frac{n_{\cdot 0} n_{\cdot 1}}{n^2}\\
    \Var_N(\hat{Y}) &= A_0 A_1 \approx \frac{n_{0 \cdot} n_{1 \cdot}}{n^2} \text{ and} \\
    \Cov_n(Y, \hat{Y}) &= \frac{n_{00} n_{11} - n_{01} n_{10}}{n^2}.
\end{align*}

\paragraph{Standard errors.}
We now turn to the PPI standard error. We have the standard deviation of the tuned PPI estimator from \cite{angelopoulos2023ppi++}
\begin{align*}
    \sigma_{\text{PPI}} &\rightarrow \sqrt{\frac{1}{n} \Var(Y - \lambda \hat{Y})} && \text{since $N \gg n$}\\
                    &= \sqrt{\frac{1}{n} \left(\Var(Y) + \lambda^2 \Var(\hat{Y}) - 2 \lambda \Cov(Y, \hat{Y}) \right)} && \text{expanding out $\Var(Y - \lambda \hat{Y})$} \\
                    &\rightarrow \sqrt{\frac{1}{n} \left( \Var(Y) - \frac{\Cov(Y, \hat{Y})^2}{\Var(\hat{Y})} \right)} && \text{plugging in $\lambda$} \\
                    &\approx \sqrt{\frac{1}{n} \left( \frac{n_{\cdot 0} n_{\cdot 1}}{n^2} - \frac{(n_{00} n_{11} - n_{01} n_{10})^2}{n^2 n_{0 \cdot} n_{1 \cdot}} \right)} && \text{plugging in $\Var(Y)$, $\Var(\hat{Y})$, $\Cov(Y, \hat{Y})$}\\
                    &= \sqrt{\frac{1}{n^2 n_{0 \cdot} n_{1 \cdot}} (n_{01} n_{00} n_{10} + n_{01} n_{00} n_{11} + n_{11} n_{10} n_{00} + n_{11} n_{10} n_{01})} && \text{using algebraic manipulation.}
\end{align*}
Above, we use the fact that as $n, N \rightarrow \infty$, the PPI tuning parameter (from Example 6.1 of \cite{angelopoulos2023ppi++}) behaves as
\begin{align*}
    \lambda \rightarrow \frac{\Cov_n(Y, \hat{Y})}{\Var_N(\hat{Y})} \approx \frac{n_{00} n_{11} - n_{01}n_{10}}{n_{0 \cdot} n_{1 \cdot}}.
\end{align*} A standard error estimator $\hat{\sigma}_{\text{PPI}}$ of $\sigma_{\text{PPI}}$ also satisfies the above equations.

We compare this with the post-stratified estimator standard error
\begin{align*}
    \hat{\sigma}_{\text{post}} &= \sqrt{A_0^2 \frac{\frac{n_{00}}{n_{0 \cdot}} \frac{n_{01}}{n_{0 \cdot}}}{n_{0 \cdot}} + A_1^2 \frac{\frac{n_{10}}{n_{1 \cdot}} \frac{n_{11}}{n_{1 \cdot}}}{n_{1 \cdot}}} \\
    &\approx \sqrt{\left( \frac{n_{0 \cdot}}{n} \right)^2 \times \frac{\frac{n_{00}}{n_{0 \cdot}} \frac{n_{01}}{n_{0 \cdot}}}{n_{0 \cdot}} + \left( \frac{n_{1 \cdot}}{n} \right)^2 \times \frac{\frac{n_{10}}{n_{1 \cdot}} \frac{n_{11}}{n_{1 \cdot}}}{n_{1 \cdot}}} && \text{plugging in $A_0$ and $A_1$}\\
    &= \sqrt{\frac{1}{n^2 n_{0 \cdot} n_{1 \cdot}} (n_{01} n_{00} n_{10} + n_{01} n_{00} n_{11} + n_{11} n_{10} n_{00} + n_{11} n_{10} n_{01})} && \text{using algebraic manipulation.}
\end{align*}
The expressions we arrive at for PPI and the post-stratified estimator agree, indicating that for large sample sizes the standard errors are approximately equal. This implies that the confidence intervals from the two methods will have the same width.

\paragraph{Point estimators.} Next, we show that the PPI and post-stratified confidence intervals are centered around the same point. Turning first to the PPI point estimator
\begin{align*}
    \hat{\theta}_{\text{PPI}} &= \lambda A_1 - \frac{1}{n} \sum_{i=1}^n (\lambda \hat{Y}_i - Y_i) \\
    &\approx \frac{\lambda n_{1 \cdot}}{n} - \frac{\lambda n_{1\cdot} - n_{\cdot 1}}{n} && \text{plugging in $A_1$, $\sum_{i=1}^n Y_i$, and $\sum_{i=1}^n \hat{Y}_i$} \\
    &= \frac{n_{\cdot 1}}{n} 
\end{align*}

The post-stratified area point estimator is
\begin{align*}
    \hat{\theta}_{\text{post}} &= A_0 \frac{n_{01}}{n_{0 \cdot}} + A_1 \frac{n_{11}}{n_{1 \cdot}} \\
    &\approx \frac{n_{0 \cdot}}{n} \times \frac{n_{01}}{n_{0 \cdot}} + \frac{n_{1 \cdot}}{n} \times \frac{n_{11}}{n_{1 \cdot}} && \text{plugging in $A_0$ and $A_1$} \\
    &= \frac{n_{\cdot 1}}{n} 
\end{align*}
We see that the final expressions are equal, which means the point estimators are approximately equal.

\end{appendices}

\end{document}